%% file: sample-sigplan.tex
\documentclass[sigplan,nonacm]{acmart}
\renewcommand\footnotetextcopyrightpermission[1]{}
\AtBeginDocument{%
  }

\usepackage{tikz}
\usepackage{amsmath}
\usepackage{listings}
\usepackage{multirow}
\usepackage[table,xcdraw]{xcolor}
\usepackage{adjustbox}
\usepackage{rotating}
\usepackage{graphicx}
\usepackage{tabularx}
\usepackage{booktabs}   
\usepackage{todonotes}
\usepackage{subfig}
\usepackage[dvipsnames]{xcolor}

\begin{document}

\title{Unifying In-Memory Data Analytics\\through Sparse Compilation}

\author{Anand Jayarajan}
\email{anandj@cs.toronto.edu}
\affiliation{%
  \institution{University of Toronto, NVIDIA}
  \country{Canada}
}

\author{Gennady Pekhimenko}
\email{pekhimenko@cs.toronto.edu}
\affiliation{%
  \institution{University of Toronto, NVIDIA}
  \country{Canada}
}

\begin{abstract}
As modern data analytics workloads become increasingly heterogeneous and hardware-intensive, achieving efficient multi-core performance across diverse applications remains an open challenge. We present Reffine, a compiler-based in-memory analytics engine that delivers high performance across a broad range of data analytics workloads. Reffine introduces a novel intermediate representation (IR), grounded in relational algebra and sparse iteration theory, that provides a unified abstraction for data and computation. This representation enables workload-agnostic, end-to-end optimizations such as operator fusion and automatic parallelization across diverse analytics applications. We further develop a sparse compiler backend that translates Reffine IR into hardware-efficient imperative code, achieving high multi-core performance without domain-specific implementations. On the TPC-H benchmark, Reffine outperforms the in-memory analytical database DuckDB by up to $24.9\times$ and the state-of-the-art compilation-based database Umbra by up to $3.2\times$. Reffine also achieves average speedups of $18.3\times$ and $47.9\times$ over Polars and NetworkX on streaming and graph analytics workloads, respectively. We plan to release Reffine as open source upon acceptance of the paper.
\end{abstract}

\settopmatter{printfolios=true}
\maketitle
\pagestyle{plain}

\input{sections/10_intro}
\input{sections/30_mot}
\input{sections/40_key}
\input{sections/60_eval}
\input{sections/70_rw}
\input{sections/80_con}

\bibliographystyle{ACM-Reference-Format}
\bibliography{acmart}

\appendix
\include{sections/90_supp}

\end{document}

%% file: sections/10_intro.tex
\section{Introduction}
Modern data analytics applications span a wide spectrum of domains, including online analytical processing (OLAP), stream processing, graph analytics, data science, scientific computing, and machine learning (ML). Achieving efficient in-memory performance across these applications often requires analytics engines to incorporate domain-specific optimizations, specialized data structures, hand-tuned implementations of common algorithms, and careful exploitation of hardware-level parallelism such as multi-core and vector execution. To meet these requirements, systems in each domain have evolved specialized execution engines with their own data and compute abstractions, internal representations, and optimization strategies. This specialization has led to a fragmented ecosystem of data analytics engines~\cite{spark,trill,storm,postgresql,pandas,duckdb}, where systems can deliver high performance within their target domains, but their abstractions and optimization techniques rarely translate to others.

Although the domain-specific data and compute abstractions help capture domain semantics and enable high-level optimizations such as query planning, algorithm selection, and domain-specific transformations, many analytics engines repeatedly address the same fundamental execution challenges such as reducing intermediate data movement, improving locality, eliminating redundant computation, and extracting parallelism on modern hardware~\cite{cidr,weld,tilt,terse,lightsaber,sparksql}. Despite solving similar underlying problems, these optimizations are often designed and implemented independently under different abstractions and execution models, resulting in substantial \emph{duplication of effort}. For example, both databases and numerical libraries optimize for locality and parallelism, yet databases do so through relational operators, columnar execution, and query pipelines, whereas numerical systems exploit multidimensional array layouts, tiling, vectorized kernels, and regular loop nests. Moreover, some systems become highly specialized, implementing optimization strategies that offer limited portability even within their own application domain. For example, modern streaming engines~\cite{grizzly,lightsaber,stream,streamboxhbm} use window-specific structure to parallelize streaming aggregations efficiently, but these techniques are tightly coupled to window semantics and do not naturally extend to other streaming operators or non-streaming workloads. Compiler-based engines alleviate some of this burden by automating transformations and code generation, but their intermediate representations are generally still designed around abstractions specific to particular workload domains~\cite{neumann,grizzly,graphmat}. Consequently, mechanisms for reasoning about locality, fusion, and parallelism remain difficult to share across otherwise different analytics systems.

As modern data processing pipelines become increasingly heterogeneous, achieving portable and generalizable performance across a broad range of application domains is no longer merely desirable, but increasingly important~\cite{weld,cidr}. Addressing this problem requires overcoming the following three key challenges. First, analytics workloads vary widely in their data modalities (e.g. tables, graphs, time-series), many exhibiting irregular access patterns and unique computation behaviors, making it hard to design a single abstraction that fits them all. Second, supporting effective optimizations that are adaptable to diverse data layouts and computation patterns is inherently complex. Third, translating these high-level computations into efficient parallel execution units, whether through multi-threading or vectorization, remains difficult without domain-specific knowledge or specialized implementation~\cite{terse,tilt,weld,split}.

To address these challenges, we propose Reffine, a novel compiler-based in-memory data analytics engine designed to achieve efficient and portable performance across a broad spectrum of applications. The key idea behind Reffine is to combine the expressive power of the well established relational model~\cite{rel} with the ability to generate efficient code for irregular data and computations using sparse compilation theory~\cite{taco}. At the core of Reffine is an intermediate representation (IR) inspired by the domain relational calculus~\cite{drc} dialect of relational algebra. Reffine IR introduces a unified data abstraction, called \emph{Field}, that maps data over a sparse, unbounded, multi-dimensional coordinate space. Next, Reffine IR provides two compute abstractions, namely \emph{Reductions} and \emph{Operators}, to define data analytics computation as a functional transformation over fields. We show that these simple yet universal abstractions give Reffine IR a highly expressive programming model capable of representing a wide variety of analytics workloads. Although Reffine IR abstractions are inherently sparse, we show that they regularize the irregular data access patterns common in analytics algorithms while explicitly capturing fine-grained data dependencies. This enables Reffine to perform end-to-end optimizations such as operator fusion and parallelization through simple IR transformations in a workload-agnostic manner. In addition, we show that Reffine IR can uncover optimization opportunities that even mature query optimizers sometimes miss (Section~\ref{sec:pass}). Finally, we build a compiler backend that translates sparse compute definitions in Reffine IR into optimized, parallel, hardware-efficient code, delivering high in-memory performance across a diverse set of workloads using a single unified abstraction.

While the goal of Reffine is to provide a unified data analytics engine, it is equally important to clarify what Reffine does not aim to do. First, Reffine is not meant to be a one-size-fits-all system~\cite{onesize} that replaces existing analytics engines. Instead, we envision Reffine as a common abstraction layer that other engines and libraries can build upon, allowing them to apply their own high-level, domain-specific optimizations (e.g., join-order optimization in SQL~\cite{job}) and then lower their computations to Reffine IR to benefit from its general-purpose optimizations and compiler-based backend. Second, although Reffine IR can express dense linear algebraic operations by using fields with zero sparsity, the current compiler backend is not optimized for such workloads. For these cases, Reffine can leverage existing tensor compilers by translating its IR as needed. Instead, in this work, we focus on data analytics workloads that lack strong compiler-based optimization and code-generation support, where Reffine provides the greatest benefit.

To demonstrate that Reffine can deliver efficient in-memory performance across a broad range of data analytics applications, we evaluate it on two benchmark suites: (1) TPC-H~\cite{tpch}, an industry-standard benchmark for decision-support workloads, and (2) a curated set of diverse analytics applications from graph analytics and streaming analytics~\cite{tilt}. We empirically measure Reffine’s performance against state-of-the-art data analytics engines and libraries such as DuckDB~\cite{duckdb}, Umbra~\cite{umbra}, NetworkX~\cite{netx}, and Polars~\cite{polars}. On a 32-core machine, Reffine outperforms DuckDB by up to $24.9\times$ and the state-of-the-art compilation-based database Umbra by up to $3.2\times$ on TPC-H. Reffine also achieves average speedups of $18.3\times$ over Polars on streaming analytics and $47.9\times$ over NetworkX on graph analytics workloads. In summary, we make the following contributions:
\begin{itemize}
    \item We highlight the limitations of the current fragmented ecosystem of data analytics engines in achieving high performance for heterogeneous, large-scale data analytics applications. To address this, we propose Reffine IR, a novel intermediate representation that can express a broad spectrum of data analytics workloads and enable effective end-to-end optimization and parallelization across them.
    \item We design and implement a compiler based on sparse compilation theory that translates Reffine IR expressions into hardware-efficient, parallelizable code. To the best of our knowledge, Reffine is the first practical system to effectively apply sparse compilation techniques to achieve high multi-core performance across a wide range of heterogeneous workloads. 
    \item We evaluate Reffine across relational, streaming, and graph analytics workloads and show that it outperforms DuckDB and Umbra by up to $24.9\times$ and $3.2\times$, respectively, on TPC-H, while achieving average speedups of $18.3\times$ over Polars on streaming analytics and $47.9\times$ over NetworkX on graph analytics.
\end{itemize}

%% file: sections/30_mot.tex
\section{Background and Motivation}\label{sec:mot}
\begin{figure*}
\footnotesize
    \begin{minipage}[t]{0.34\textwidth} 
        \begin{lstlisting}[numbers=none, captionpos=b, language=SQL]
SELECT pk, sk, qty FROM PartSupp
WHERE pk IN (SELECT pk FROM Part
  WHERE size > 10) AS V1
AND sk IN (SELECT sk FROM Supplier
  WHERE nation = "US") AS V2
ORDER BY pk, sk;
        \end{lstlisting}
    \end{minipage}%
    \begin{minipage}[t]{0.32\textwidth}
        \begin{lstlisting}[numbers=none, captionpos=b]
def pagerank_step(M, w, N):
    M_hat = 0.85 * M
    v = M_hat @ w + (0.15 / N)
    return v
        \end{lstlisting}
    \end{minipage}%
    \begin{minipage}[t]{0.32\textwidth}
        \begin{lstlisting}[numbers=none, captionpos=b, mathescape]
dma50 = Stock.Window(50, 1)
    .Sum(e -> e.price)
    .Select(e -> e/50);
out = Stock.Join(dma50,
    (e, avg) $\rightarrow$ e.price > avg);
        \end{lstlisting}
    \end{minipage}
    \vspace{-10pt}
    \caption{(a) Simplified TPC-H Query, (b) one step of PageRank algorithm, (c) $50$-day moving average on a stock price stream}
    \label{fig:eg}
\end{figure*}

\begin{figure*}[th]
    \footnotesize
    \begin{minipage}[t]{0.31\textwidth} 
        \begin{equation*}
            \begin{aligned}
                V1 =& \forall pk : [Part[pk] \neq \phi \land \\
                    & Part[pk].size > 10] \{\}\\
                V2 =& \forall sk : [Supplier[sk] \neq \phi \land \\
                    & Supplier[sk].nation = \text{"US"}] \{\}\\
                Out =& \forall pk, sk : [V1[pk] \neq \phi \land V2[sk] \neq \phi \land \\
                     & PartSupp[pk, sk] \neq \phi] \{ PartSupp[pk, sk].qty \}
            \end{aligned}
        \end{equation*}
    \end{minipage}%
    \begin{minipage}[t]{0.34\textwidth}
        \begin{equation*}
            \begin{aligned}
                M\_hat =& \forall x, y : [M[x,y] \neq \phi] \{ M[x,y] * 0.85 \}\\
                v =& \forall x : [M\_hat[x] \neq \phi] \{\\
                &d = \forall y : [M\_hat[x,y] \neq \phi \land w[y] \neq \phi]\\
                    & \qquad \{ M[x,y] * w[y] \} \\
                    & \oplus(\text{SUM}, d) + (0.15 / N)\\
                    & \}
            \end{aligned}
        \end{equation*}
    \end{minipage}%
    \begin{minipage}[t]{0.33\textwidth}
        \begin{equation*}
            \begin{aligned}
                wsum =& \forall t : [Stock[t-50: t] \neq \phi]\\
                \{ & \oplus(\text{WSUM}, Stock[t-50: t])/50 \}\\
                dma =& \forall t : [wsum[t] \neq \phi] \{ wsum[t]/50 \}\\
                out =& \forall t : [Stock[t] \neq \phi \land dma50[t] \neq \phi]\\
                     & \{ Stock[t] > dma50[t] \}
            \end{aligned}
        \end{equation*}
    \end{minipage}
    \vspace{-5pt}
    \caption{Reffine IR definition of the example (a) TPC-H Query, (b) PageRank algorithm, (c) $50$-day moving average}
    \vspace{-5pt}
    \label{fig:egir}
\end{figure*}

In this section, we illustrate the key principles that guided the design of Reffine using three representative applications taken from distinct analytics domains, as shown in Figure~\ref{fig:eg}. The first is a simplified version of Query 20 from the industry-standard decision support benchmark TPC-H~\cite{tpch}. This query is written on three tables \texttt{Part}, \texttt{Supplier}, and \texttt{PartSupp} with $pk$, $sk$, and $(pk, sk)$ as the primary keys respectively. The query selects rows in \texttt{PartSupp} that has parts with size greater than $10$ and has suppliers in US. The second is a temporal query that computes a $50$-day moving average over a stock price time-series data and compare it against the current price. The final one is a single step of the PageRank algorithm written in NumPy~\cite{numpy} with the graph implemented using an adjacency matrix.

\noindent\textbf{Optimizing across operation boundaries.}
Many analytics engines execute applications as compositions of primitive operations, passing tuples or batches between them. While modular, this execution model can incur substantial overhead from intermediate materialization and repeated data movement across operator boundaries. Fusion mitigates this overhead by combining dependent operations and keeping intermediate values local. In practice, however, fusion is commonly implemented through transformation rules that recognize particular operator patterns and rewrite them into more efficient forms. Such rules can be effective for known patterns, but they are inherently tied to the abstractions and semantics of the corresponding domain.

For example, DuckDB evaluates the query in Figure~\ref{fig:eg}a with two nested subqueries independently before processing the outer query. A more efficient way is to rewrite the query as a three-way join (Listing~\ref{lst:tcphopt}) that co-iterates over \texttt{Part}, \texttt{Supplier}, and \texttt{PartSupp}, applying the predicates inline and avoiding the materialization of the intermediate subquery results. Achieving this transformation requires the optimizer to recognize the relationship between the nested subqueries and the outer query and rewrite them into an equivalent join structure. Although such transformations can be encoded as query-rewrite rules, prior work has shown that pattern-based optimization rules often generalize poorly: small changes in query structure can prevent otherwise applicable transformations from firing~\cite{optbig}.

\begin{figure}[h]
\footnotesize
\begin{lstlisting}[label={lst:tcphopt}, numbers=none, captionpos=b, caption={Optimized example TPC-H query}]
    SELECT PS.pk, PS.sk, PS.qty
    FROM PartSupp PS, Part P, Supplier S
    WHERE PS.pk = P.pk AND PS.sk = S.sk
      AND P.size > 10 AND S.nation = "US"
    ORDER BY PS.pk, PS.sk;
\end{lstlisting}
\vspace{-10pt}
\end{figure}

A similar optimization opportunity appears in the temporal query in Figure~\ref{fig:eg}c, but realizing it requires an entirely different set of domain-specific transformations. The query first computes a sliding-window sum, converts it to an average, and then joins the result back with the input stream. A more efficient implementation maintains the running sum and performs the final comparison directly inside the stateful window aggregation, eliminating both the intermediate stream and the temporal join. However, deriving this form requires reasoning about window semantics, state updates, and temporal alignment—semantics that are completely different from the above transformation. Existing streaming systems may fuse simple producer-consumer operator sequences, but automatically synthesizing this stateful fused aggregation remains beyond the capabilities of current engines.

These examples expose a common limitation: the same optimization principle is implemented through different domain-specific rewrite rules, which are often fragile and miss opportunities as computation patterns become more complex. We observe that, for a broad class of fusion transformations, the optimizer need not reason about the domain-specific semantics of every individual operator. Instead, it is sufficient to understand the fine-grained data dependencies between operations and to express their combined computation. This observation leads to three design principles: (1) a data representation that explicitly exposes fine-grained data dependencies, (2) a flexible and expressive compute representation capable of capturing the combined semantics of fused operations, and (3) general transformation rules that operate on the information exposed by these representations rather than on domain-specific operator patterns.

\noindent\textbf{Efficient execution over irregular data.}
Data analytics workloads frequently exhibit highly irregular data access patterns, such as data shuffling, dictionary lookups, and random indexing, that complicate efficient in-memory execution. Graph analytics is a canonical example, as traversing large-scale graphs (e.g., social networks) requires pointer chasing, resulting in unpredictable memory accesses and poor cache locality~\cite{graphmat}. One potential solution is to remap irregular data structures into dense representations to obtain regularity. For example, Figure~\ref{fig:eg}b shows a PageRank iteration implemented in NumPy using a dense adjacency matrix. Although this design enables predictable memory access patterns that align well with modern hardware, it introduces substantial redundant computation as real-world graphs are extremely sparse.

To efficiently handle irregular data and computation patterns, one promising approach is to use sparse data structures and specialized kernels tailored for them. Recent works~\cite{taco,indexstream} have shown that many common data structures and operations used in relational databases and graph analytics can be expressed as linear algebraic operations over sparse tensors. For example, the graph can be represented using a sparse adjacency matrix and use efficient sparse matrix-vector multiplication kernels in numerical libraries like SciPy~\cite{scipy} to perform the page ranking step. However, expressing analytics computations in sparse form typically requires substantial manual effort or relies on highly specialized, hand-engineered implementations. Although prior works~\cite{indexstream,graphmat} have explored automatically mapping relational queries and graph analytics applications onto sparse operations, these approaches are often restricted to a narrow class of operations, and cannot generalize across the full spectrum of modern data analytics applications. This gap highlights the need for a compiler-based approach that can automatically map high-level, irregular computations onto efficient sparse representations and generate hardware-efficient code.

\noindent\textbf{Extracting parallelism.}
Data analytics engines exploit hardware parallelism in different ways. Systems such as ClickHouse~\cite{clickhouse} and NumPy~\cite{numpy} use hand-tuned primitives to leverage SIMD~\cite{avx}, while engines such as Dask~\cite{dask}, Trill, and DuckDB parallelize individual operators or over explicitly partitioned inputs. Effectively exploiting hardware parallelism requires understanding the data dependencies and access boundaries of the queries. This becomes difficult when faced with complex operator compositions that involve non-trivial data dependencies~\cite{tilt}.

The temporal query shown in Figure~\ref{fig:eg}c uses a sliding-window sum, which is typically implemented using a stateful aggregation function in streaming analytics engines like Trill~\cite{trill} to avoid redundant computation over overlapping events. This, however, introduces a sequential dependency between consecutive events. Furthermore, merging the output of the window operation back into the input stream via a temporal join introduces yet another self-dependency on the input stream. In such cases, we observe that many analytics engines fail to parallelize the query end-to-end and instead fall back to sequential execution~\cite{tilt}. We argue that effective parallelization requires fine-grained dependency information to identify the input regions contributing to each output and where execution can be safely partitioned.

\noindent\textbf{Key insights.}
Based on the above observations, we derive the following design principles for Reffine as the foundation of a unified data analytics engine. First, we need a data representation based on sparse abstractions that can express a wide range of data modalities and uniformly capture both regular and irregular structures. Second, we need an expressive compute representation that can support the diverse computation patterns found in data analytics, is flexible enough to compose into semantically consistent fused operators, and exposes fine-grained data dependencies within the computation. Finally, a compiler backend that lowers these logical abstractions into efficient physical implementations on modern hardware, extensively exploiting multi-core parallelism and leveraging SIMD optimizations whenever possible.

%% file: sections/40_key.tex
\section{Reffine: A Unified Data Analytics Engine}
Reffine consists of two main components: (1) an intermediate representation called Reffine IR, and (2) a compiler-based execution engine. Reffine IR introduces a data abstraction called \emph{Fields}, which arranges data points in a sparse, unbounded multi-dimensional coordinate space and can model a wide variety of data structures used in modern analytics applications. Next, Reffine IR introduces two simple constructs, namely \emph{Operators} and \emph{Reductions}, for defining computations as a functional transformation on top of fields. Reffine expressions are grounded in first-order logic, providing high expressive power for defining a wide range of computations.

The combination of Reffine’s data and compute abstractions offers several key benefits. First, expressing data as fields regularizes irregular data patterns and computations, making them easier to analyze and reason about. On top of that, Reffine Operators and Reductions expose fine-grained dependencies between input and output fields, enabling the compiler to identify inherent data parallelism and perform optimizations such as operator fusion through simple and generalizable IR transformations. Finally, we develop a compiler backend that automatically lowers sparse compute definitions in Reffine IR into hardware-efficient parallel loops using sparse compilation techniques.

Although Reffine can operate as a standalone engine, we find that it is most effective when integrated with existing data analytics engines such as relational databases, stream processing systems, and numerical libraries. These front-end systems provide user-friendly domain-specific languages (DSLs) and rich domain knowledge that enable high-level, domain-specific optimizations. After such optimizations are applied, the resulting computation can be lowered into Reffine IR, which is expressive enough to directly represent many primitive operations from existing DSLs (Section~\ref{sec:ir}). Reffine then applies a sequence of IR transformations to fuse and parallelize the computation (Section~\ref{sec:pass}). Finally, it leverages SMT solvers such as Z3 to analyze the first-order-logic-based Reffine IR expressions, synthesize corresponding imperative loops into LLVM IR, and ultimately generate executable code (Section~\ref{sec:codegen}). The following sections describe each stage of this lifecycle in detail.

\subsection{Reffine IR}\label{sec:ir}
\noindent\textbf{Fields} are infinitely wide coordinate spaces with an arbitrary number of dimensions, where each coordinate is associated with a value, represented as $\{ \langle \text{coordinate} \rangle, \text{value} \}$. The $coordinate$ is a tuple of integers\footnote{non-integer data types like strings can also be supported by defining a hashing function to map them to unique integers} and $value$ is the value associated with that coordinate and can be of any type like integers, floating points, arrays, structures, or even Fields. 

Reffine uses this single abstraction for representing all types of data structures that it supports. To give a few examples, scalar values can be represented using zero dimensional fields. Time-series data streams can be modeled as one dimensional fields, where timestamps serve as coordinates and event payloads as the associated values. Graphs can be represented as two dimensional fields, with source and destination nodes as coordinates and edge weights as values. Relational tables can similarly be expressed by using the primary key attributes as coordinates and the remaining attributes as the value. For coordinate points where the underlying dataset doesn't have a corresponding value, Reffine assumes a null value ($\phi$) to indicate the absence of a value at that position.

The Field abstraction provides several key properties. First, every data point is uniquely identified by its coordinate, enabling Reffine to track data dependencies at a fine-grained level. Second, the infinitely wide and sparse representation makes irregular data structures appear more regular, simplifying analysis and optimizations. Finally, fields impose a well-defined ordering over values based on their coordinates, making iteration over data points more predictable, especially when co-iterating over multiple fields.

Reffine IR supports a few ways to access data points in a field. For example, assume PartSupp table in Figure~\ref{fig:eg}a can be represented as a 2D field $\{<pk, sk>, qty\}$. The value of PartSupp corresponding to $pk = 123$ and $sk = 456$ can be referred using the expression $PartSupp[123, 456]$. Reffine also allows partial indexing into the field. For example, $PartSupp[123]$ refers to a one-dimensional sub-field ($\{<sk>, qty\}$) with all the data points in PartSupp with $pk=123$. Reffine also supports range-based indexing. For example, $PartSupp[100:200]$ represents the slice of PartSupp with $100 < pk \leq 200$.

\noindent\textbf{Reduction} in Reffine IR, denoted as $\oplus \left(f, F\right)$ where $f$ is a reduction function and $F$ is a field, is used to convert fields into scalar values. Reduction functions are defined by two properties: (1) an $init$ function that initializes a state, (2) an $acc$ function that accumulates the values in the field in serial order into the state, and (3) an optional $deacc$ function that defines the inverse of $acc$. Reffine, by default, supports common aggregation functions such as SUM, COUNT, MIN, MAX and many more. For example, COUNT reduction function is defined as follows:

\vspace{-10pt}
\begin{equation*}
\small
\begin{aligned}
COUNT: \{init() \rightarrow 0;\quad acc(state, val) \rightarrow state + 1;\}
\end{aligned}
\end{equation*}

More complex aggregations can be defined by combining existing ones or by writing new reduction functions.

\noindent\textbf{Operator} defines a new field as a functional transformation over one or more input fields. A Reffine operator has three components as shown below:

\vspace{-7pt}
\begin{equation*}
\begin{aligned}
Out = \forall iterators: [predicate] \{ value \}
\end{aligned}
\end{equation*}

The \emph{iterators} define one or more symbols representing the coordinates of the output field. The \emph{predicate} is a first-order logical expression, and the \emph{value} is another Reffine IR expression. For each coordinate defined by the iterators, the output field assumes the corresponding value when the predicate evaluates to $true$. We design Reffine operators as an extension of domain relational calculus~\cite{drc}. As a result, Reffine inherits the full expressive power of relational algebra. For example, a simple SQL query like \textit{(SELECT pk, SUM(qty) FROM PartSupp WHERE pk < 50 GROUP BY pk)} can be expressed in Reffine IR as follows:

\begin{equation*}
\small
\begin{aligned}
Out =& \forall pk: [PartSupp[pk] \neq \phi \And pk < 50]\\
&\{ \quad \oplus(SUM, PartSupp[pk]) \quad \}
\end{aligned}
\end{equation*}

In this expression, the output is a one-dimensional field with $pk$ as the symbolic iterator. For each coordinate point, the output field assumes the value given by the sum of the values of the sub-field $PartSupp[pk]$, provided that the sub-field exists (i.e., not null) and $pk$ is less than $50$. Next, we demonstrate that Reffine can represent complex and diverse computations. Figure~\ref{fig:egir} shows the direct translation of all three example applications in Figure~\ref{fig:eg} to Reffine IR.

The Reffine IR expression in Figure~\ref{fig:egir}a defines two intermediate fields, $V1$ and $V2$, corresponding to the inner queries in the SQL version. The output field $Out$ is then defined over the coordinates $(pk, sk)$, along with a predicate. When $PartSupp$ has a non-null value at $(pk, sk)$, and both $V1$ and $V2$ have non-null values at $pk$ and $sk$, respectively, the $Out$ field assumes the value of the qty attribute of PartSupp at the same coordinate. Similarly, in Figure~\ref{fig:egir}b, $M\_hat$ is defined as an element-wise multiplication over the input two-dimensional field $M$, and $v$ defines a matrix-vector multiplication plus element addition over $M\_hat$ and one-dimensional $w$ field. Although not our primary focus, Reffine IR can similarly express dense linear algebra operations as well. However, we defer efficient support for such operations to future work.

Finally, Figure~\ref{fig:egir}c shows the Reffine IR version of the $50$-day moving average query. The predicate $Stock[t-50: t] \neq \phi$ defines a $50$-day sliding window on the input one-dimensional $Stock$ field, and $\oplus(\text{WSUM}, Stock[t-50: t])$ sums up the values within the $(t-50: t]$ slice of the $Stock$. WSUM is a reduction function defined in Reffine to efficiently compute moving window sums using $acc$ and $deacc$ functions without side-effects.

The above examples demonstrate that the three abstractions, fields, operators, and reductions, make Reffine IR a highly expressive programming model capable of representing a broad spectrum of data analytics applications.

\subsection{IR Transformations}\label{sec:pass}
Once a data analytics application is lowered to Reffine IR, it undergoes a series of IR transformations to produce a semantically equivalent but more efficient compute definition. Although we believe Reffine IR can support a wide range of sophisticated optimizations, a full exploration of such opportunities is beyond the scope of this work. Instead, we focus on two key optimizations that offer the greatest impact, namely operator fusion and parallelization.

\noindent\textbf{Operator fusion}
Reffine achieves operator fusion by iteratively replacing references to intermediate fields with their defining expressions. For example, following is the final operator in the example relational query in Figure~\ref{fig:egir}a.

\vspace{-10pt}
\begin{equation*}
\small
\begin{aligned}
Out =& \forall pk, sk: [V1[pk] \neq \phi \land V2[sk] \neq \phi \land \\
     & PartSupp[pk, sk] \neq \phi] \{ PartSupp[pk, sk].qty \}
\end{aligned}
\end{equation*}

In this operator, $V1[pk] \neq \phi$ checks whether the field $V1$ contains a non-null value at coordinate $pk$. Since the existence of $V1[pk]$ is determined entirely by the predicate in the operator that defines $V1$, which is $Part[pk] \neq \phi \land Part[pk].size > 10$, Reffine can replace $V1[pk] \neq \phi$ expression with this predicate directly. A similar substitution can be applied to $V2[sk]$ as well. Reffine applies such transformations repeatedly and greedily until no further fusion is possible. The final fused expression for $Out$ is as follows:

\vspace{-10pt}
\begin{equation}
\small
\begin{aligned}
Out =& \forall pk, sk: [Part[pk] \neq \phi \land Part[pk].size > 10 \land \\
& Supplier[sk] \neq \phi \land Supplier[sk].nation = \text{"US"} \land\\
& PartSupp[pk, sk] \neq \phi] \{ PartSupp[pk, sk].qty \}
\end{aligned}
\label{eqn:fusesql}
\end{equation}

Through fusion, the entire SQL query in Figure~\ref{fig:eg}a is expressed as a single Reffine IR operator. Moreover, this fused version is identical to the result we would obtain by lowering the optimized SQL query in Listing~\ref{lst:tcphopt}. This demonstrates that simple and generalizable transformations such as operator fusion in Reffine can realize optimization opportunities that would otherwise require specialized rule-based transformations at higher abstraction levels like SQL. Similarly, the fused versions of the streaming query is as follows:
 
\begin{equation}
\small
\begin{aligned}
Out =& \forall t : [Stock[t-50: t] \neq \phi] \\
     & \quad \{ Stock[t] > \oplus(\text{WSUM}, Stock[t-50: t])/50 \}
\end{aligned}
\label{eqn:fusedma}
\end{equation}

Although Reffine makes operator fusion simple and straightforward, there are cases where operators cannot be fused. For instance, when the coordinates are reordered or new coordinates are derived from the values. In such cases, Reffine needs to materialize the output before proceeding to the next set of operators. That said, our greedy approach to fusion ensures all the fusible operators can be fused together and need to materialize only necessary intermediate results.

\noindent\textbf{Operator parallelization}
Since Reffine operators define how each output coordinate is evaluated, our parallelization strategy is to modify the operator so that it produces the output field only within a specified boundary and then parallelize the operator across disjoint boundaries. This is achieved by adding symbolic boundary conditions to the predicate expression of the operator, constraining the coordinate iterators to a particular slice of the output field. For example, to restrict the operator in Equation~\ref{eqn:fusedma} to produce output only for the interval $[start, end)$, we can add $t \geq start \land t < end$ into the operator as shown below:

\vspace{-10pt}
\begin{equation*}
\footnotesize
\begin{aligned}
Out =& \forall t : [Stock[t-50: t] \neq \phi \land t \geq start \land t < end]\\
&\quad \{ Stock[t] > \oplus(\text{WSUM}, Stock[t-50: t])/50 \}
\end{aligned}
\end{equation*}

Once these additional constraints are added to the predicate, parallelization simply reduces to invoking the operator with different non-overlapping $[start, end)$ ranges. Reffine’s side-effect–free compute definitions allow us to apply such parallelization strategies without violating the semantics of the operator or data dependencies. However, we observe that slicing the output field along multiple dimensions often complicate the generated code and can even degrade performance. Hence, we adopt a simpler strategy that partitions only along the outermost coordinate.

\noindent\textbf{Canonicalization}
After the fusion and parallelization passes, the final step is to canonicalize the Reffine IR expression. This stage applies standard compiler optimizations such as common subexpression elimination and dead code elimination. At this stage, Reffine also converts range based predicates like $Stock[t-50: t] \neq \phi$ to its semantically equivalent form $(Stock[t] \neq \phi \lor Stock[t-50] \neq \phi)$. In addition, every Reffine IR operator is rewritten into a normalized single-iterator form, as shown below:

\vspace{-10pt}
\begin{equation}
\footnotesize
\begin{aligned}
Out &= \forall pk : [PartSupp[pk] \neq \phi \land Part[pk] \neq \phi \land \\
    & Part[pk].size > 10 \land pk \geq start \land pk < end] \{ \\
    & \quad \forall sk : [PartSupp[pk][sk] \neq \phi \land Supplier[sk] \neq \phi \land \\
    & \quad Supplier[sk].nation = \text{"US"}] \{PartSupp[pk][sk].qty\}\\
\}
\end{aligned}
\label{eqn:cansql}
\end{equation}

Operators that originally contain multiple iterators are converted into a nested structure like the above by distributing the iterator predicates to the outer operators and inner operators for $pk$ and $sk$ iterators respectively. This transformation places all operators into a uniform, simplified canonical form prior to code generation.

\subsection{Solver-Assisted Code Generation}\label{sec:codegen}

\begin{table*}[ht]
\centering
\begin{tabular}{|l|ll|}
\hline
\rowcolor[HTML]{C0C0C0} 
\textbf{Iteration space}                                                                & \multicolumn{2}{l|}{\cellcolor[HTML]{C0C0C0}\textbf{Properties}}                                                                                                                                                                                                                                                                                                                                                                                                                                                                                                                                                                                                                                                                                                                                                                                             \\ \hline
\begin{tabular}[c]{@{}l@{}}UniversalSpace\\ $\forall t$\end{tabular}                    & \multicolumn{2}{l|}{\begin{tabular}[c]{@{}l@{}}$init$: $-\infty$; $\quad cond(idx)$: $true$; $\quad is\_alive(idx)$: $true$;\\  $to\_iter(idx)$: $idx$; $\quad to\_idx(iter)$: $iter$; $\quad next(idx)$: $idx+1$;\end{tabular}}                                                                                                                                                                                                                                                                                                                                                                                                                                                                                                                                                                                                                             \\ \hline
\begin{tabular}[c]{@{}l@{}}ConstantSpace\\ $\forall t: [t==k]$\end{tabular}             & \multicolumn{2}{l|}{\begin{tabular}[c]{@{}l@{}}$init$: $k$; $\quad cond(idx)$: $to\_iter(idx) == k$; $\quad is\_alive(idx)$: $true$;\\ $to\_iter(idx)$: $idx$; $\quad  to\_idx(iter)$: $iter$; $\quad next(idx)$: $idx$;\end{tabular}}                                                                                                                                                                                                                                                                                                                                                                                                                                                                                                                                                                                                                       \\ \hline
\begin{tabular}[c]{@{}l@{}}CoordSpace\\ $\forall t:[F[t] \neq \phi]$\end{tabular}       & \multicolumn{2}{l|}{\begin{tabular}[c]{@{}l@{}}$init$: $F[0]$; $\quad cond(idx)$: $(idx < len(F) \And F.bitmap[idx])$; $\quad is\_alive(idx)$: $idx < len(F)$;\\ $to\_iter(idx)$: $F[idx]$; $\quad to\_idx(iter)$: $locate(F, iter)$; $\quad next(idx)$: $idx+1$;\end{tabular}}                                                                                                                                                                                                                                                                                                                                                                                                                                                                                                                                                                              \\ \hline
\begin{tabular}[c]{@{}l@{}}LBoundSpace\\ ($\forall t: [t\geq lb]$)\end{tabular}         & \multicolumn{2}{l|}{\begin{tabular}[c]{@{}l@{}}$init$: $min(base.init, lb)$; $\quad cond(idx)$: $base.cond(idx) \And to\_iter(idx) \geq lb$;\end{tabular}}                                                                                                                                                                                                                                                                                                                                                                                                                                                                                                                         \\ \hline
\begin{tabular}[c]{@{}l@{}}UBoundSpace\\ $\forall t: [t \leq ub]$\end{tabular}          & \multicolumn{2}{l|}{\begin{tabular}[c]{@{}l@{}}$cond(idx)$: $base.cond(idx) \And to\_iter(idx) \leq ub$;\\ $is\_alive(idx)$: $base.is\_alive(idx) \And to\_iter(idx) \leq ub$;\end{tabular}}                                                                                                                                                                                                                                                                                                                                                                                                                                                                                                           \\ \hline
\begin{tabular}[c]{@{}l@{}}ShiftedSpace\\ ($\forall t:[F[t+k] \neq \phi]$)\end{tabular} & \multicolumn{2}{l|}{\begin{tabular}[c]{@{}l@{}}$init$: $base.init + k$; $\quad to\_iter(idx)$: $base.to\_iter(idx) + k$; $\quad to\_idx(iter)$: $base.to\_idx(iter - k)$;\end{tabular}}                                                                                                                                                                                                                                                                                                                                                                                                                                                                                                                                                 \\ \hline
\begin{tabular}[c]{@{}l@{}}FilterSpace\\ ($\forall t : [expr]$)\end{tabular}            & \multicolumn{2}{l|}{\begin{tabular}[c]{@{}l@{}}$cond(idx)$: $base.cond(idx) \And expr$;\end{tabular}}                                                                                                                                                                                                                                                                                                                                                                                                                                                                                                                                                   \\ \hline
\begin{tabular}[c]{@{}l@{}}InterSpace\\ ($\forall t : [left \land right]$)\end{tabular}  & \multicolumn{1}{l|}{\begin{tabular}[c]{@{}l@{}}$init$: $min(left.init, right.init)$;\\ $is\_alive(idx)$: $left.is\_alive(idx[0]) \And right.is\_alive(idx[1])$;\\ $to\_iter(idx)$: $min(left.to\_iter(idx[0]), right.to\_iter(idx[1]))$;\\ $to\_idx(iter)$: $\{left.to\_idx(iter), right.to\_idx(iter)\}$;\\ $cond(idx)$:\\ $\quad liter = left.to\_iter(idx[0])$\\ $\quad riter = right.to\_iter(idx[1])$;\\$\quad newcond = left.cond(idx[0]) \And right.cond(idx[1]);$\\ $\quad$return $(newcond \And  liter == riter)$;\end{tabular}} & \begin{tabular}[c]{@{}l@{}}next(dx):\\ $\quad liter = left.to\_iter(idx)$;\\ $\quad riter = right.to\_iter(idx)$;\\ $\quad lidx = left.next(idx[0])$;\\ $\quad ridx = right.next(idx[1])$;\\ $\quad$return $\{$\\ $\qquad (liter \leq riter)$ ? $lidx$ : $idx[0],$\\ $\qquad (riter \leq liter)$ ? $ridx$ : $idx[1]$\\ $\quad\}$\end{tabular} \\ \hline
\begin{tabular}[c]{@{}l@{}}UnionSpace\\ ($\forall t : [left \lor right]$)\end{tabular}    & \multicolumn{2}{l|}{\begin{tabular}[c]{@{}l@{}}$init$: $max(left.init, right.init)$; $\quad cond(idx)$: $(left.cond(idx[0]) || right.cond(idx[1]))$;\\ $to\_iter(idx)$: $min(left.to\_iter(idx[0]), right.to\_iter(idx[1])$;\\ $ to\_idx(iter): \{left.to\_idx(iter), right.to\_idx(iter)\}$;\\ $is\_alive(idx)$: $left.is\_alive(idx[0]) || right.is\_alive(idx[1])$; $\quad next(idx)$: same as InterSpace\end{tabular}}                                                                                                                                                                                                                                                                                                                                                                                                        \\ \hline
\end{tabular}
\caption{Predefined iteration spaces in Reffine and their corresponding properties (same as that of the base if unspecified)}
\label{tbl:iterspace}
\vspace{-10pt}
\end{table*}

After IR transformations, the next step is to generate imperative loops corresponding to each operator. A na\"ive way to convert any Reffine IR operator of the form $Out = \forall iter : [pred] \{ value \} $ into a loop would be to increment $iter$ from $-\infty$ to $\infty$ and write $(iter, value)$ pair to the output if $pred$ evaluates to true. This approach is clearly infeasible, as it performs an enormous amount of redundant work over an unbounded sparse iteration space. Instead, Reffine constructs the loops by pruning, or as we call it \emph{refining}, the iteration space to only iterate through coordinate points that evaluates $pred$ to $true$. For this, Reffine finds a mapping between the sparse logical iterator $iter$ into a compact physical index $idx$ that ranges only over coordinates that are most likely to produce a non-null value, yielding far more efficient loops.

Reffine constructs index-based iteration spaces by analyzing the operator’s first-order–logic predicate $pred$ using an SMT solver such as Z3~\cite{z3}. Reffine formally defines an iteration space using six properties: (1) $init()$ function as the starting point of the iterator, (2) $is\_alive(idx)$ a boolean function to check if the iteration has finished, (3) $to\_idx(iter)$ function to convert a given iterator to corresponding index, (4) $to\_iter(idx)$ to convert an index to the corresponding iterator, (5) $cond(idx)$ a boolean function to check if an iteration is non-null, (6) $next(idx)$ function to return the next index. Based on these properties, Reffine supports a set of predefined iteration spaces as shown in Table~\ref{tbl:iterspace}. Iteration spaces, such as UniversalSpace, ConstantSpace, and CoordSpace, are base iteration spaces, where UniversalSpace defines the default iteration domain from $-\infty$ to $\infty$. CoordSpace iterates only over the non-null coordinates along a particular axis of a materialized field. The remaining iteration spaces are derived from other spaces. For example, LBoundSpace and UBoundSpace apply lower and upper bounds to a base space, respectively. InterSpace and UnionSpace construct intersections and unions of existing iteration spaces.

\begin{figure}[]
    \centering
    \begin{minipage}{0.8\linewidth}
        \subfloat{
            \includegraphics[width=0.5\textwidth]{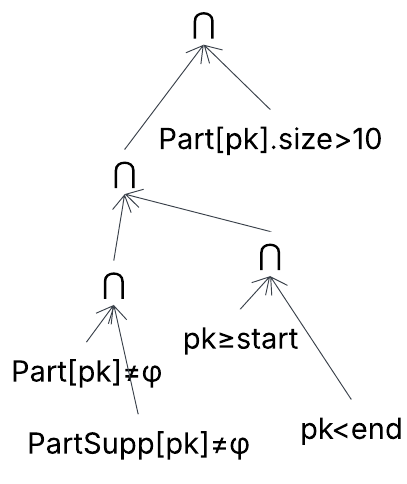}
            \label{fig:unopt_iter}
        }%
        \subfloat{
            \includegraphics[width=0.5\textwidth]{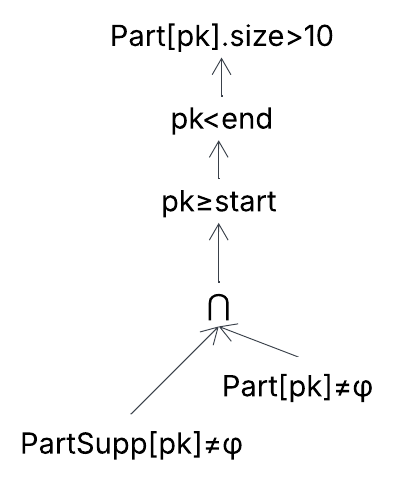}
            \label{fig:opt_iter}
        }
    \end{minipage}
    \vspace{-5pt}
    \caption{Iteration spaces: (a) Initial, (b) Simplified}
    \vspace{-10pt}
\end{figure}

For any given combination of $iter$ and $pred$, Reffine constructs the corresponding iteration space by composing these predefined iteration spaces. To do this, Reffine first decomposes $pred$ expression into its constituent primitive expressions each connected using $\land$ or $\lor$. Each constituent expression is then analyzed using the Z3 solver to find the best matching predefined iteration space listed in Table~\ref{tbl:iterspace}. After that, Reffine builds the iteration space for the full predicate by combining these spaces by applying InterSpace ($\land$) or UnionSpace ($\lor$) as shown in Figure~\ref{fig:unopt_iter}.

To take a concrete example, consider the outer operator in the Reffine IR expression shown in Equation~\ref{eqn:cansql}. The predicate over the iterator $pk$ can be broken down to $PartSupp[pk] \neq \phi$, $Part[pk] \neq \phi$, $pk \geq start$, $pk < end$ and $Part[pk].size > 10$ all combined together with $\land$. Reffine maps each of these expressions to a concrete iteration space definition. However, since boolean expressions can appear in many equivalent forms, it makes direct pattern matching unreliable. To robustly determine the correct iteration space, Reffine relies on Z3 solver. To do this, Reffine uses a parameterized template expression ($template$) for each iteration space in Table~\ref{tbl:iterspace}. To determine whether a constituent predicate expression ($expr$) matches an iteration space, Reffine checks satisfiability of $\forall iter : (expr == template)$. If satisfiable, the expression is mapped to that iteration space.

For instance, for mapping the upper bound condition $pk < end$ to UBoundSpace, Reffine uses the template $pk \leq A*start + B*end + C$, a weighted sum of symbolic constants ($A*start + B*end$) in the Reffine operator and a bias term ($C$). Then Reffine solves for $\forall pk : ((pk < end) == (pk \leq A*start + B*end + C)$ using Z3. Since this is satisfiable, Reffine maps the expression to UBoundSpace using the parameters returned by Z3, yielding a boundary of $end - 1$ in this case. If unsatisfiable, Reffine proceeds to match the next template. If none of the templates match, Reffine defaults to constructing a FilteredSpace over the expression. For example, conditions over the values like $Part[pk].size > 10$ doesn't satisfy any templates and therefore is mapped to the FilteredSpace. Once all the constituent expressions are mapped to a predefined iteration space, Reffine merges them using InterSpace to construct the iteration space for the full predicate.

Figure~\ref{fig:unopt_iter} shows the initial iteration-space generated for the outer operator in Equation~\ref{eqn:cansql}. However, directly lowering this iteration space into a loop is often inefficient, as it contains several InterSpaces. Since InterSpaces and UnionSpaces co-iterate over multiple fields, the generated code for them tends to be more complex, requiring conditional checks and the evaluation of long expressions, as shown in Table~\ref{tbl:iterspace}. However, in many cases, InterSpaces between derived iteration spaces can be simplified. For example, an intersection between LBoundSpace and UBoundSpace can be rewritten as an LBoundSpace derived from a UBoundSpace, or vice versa. These rewrites help eliminate unnecessary InterSpace nodes and significantly simplify the iteration-space as shown in Figure~\ref{fig:opt_iter}. Here, the only necessary InterSpace is between the two CoordSpaces of $PartSupp$ and $Part$.

Once the iteration space is prepared, the loop is generated using the template shown in Figure~\ref{fig:loop}a. The loop initializes $idx$ using $to\_idx(init())$ expression and then iterates through by incrementing the index $idx$ using $next(idx)$ until $is\_alive(idx)$ becomes false. For each iteration that passes the $cond(idx)$, Reffine evaluates the $iter$ and the $value$ and writes to the output. For co-iterating over multiple fields, Reffine employs either merge-based or hash-based strategies to efficiently traverse the inputs~\cite{taco}. Similarly, for reductions, the state is initialized prior to the loop, and the value is accumulated based on the reduction function.

\subsection{Execution Engine}

\begin{figure}
\raggedright
\begin{minipage}{0.4\linewidth}
\footnotesize
\begin{lstlisting}[numbers=none, captionpos=b, mathescape, language=c]
i = to_idx(init());
while(is_alive(i)) {
  if (cond(i)) {
    iter=to_iter(i);
    out$\leftarrow$(iter,value);
  }
  i = next(i);
}
\end{lstlisting}
\end{minipage}\hspace{12pt}
\begin{minipage}{0.4\linewidth}
\footnotesize
\begin{lstlisting}[numbers=none, captionpos=b, language=c]
i = to_idx(init());
while(is_alive(i)) {
  iter = to_iter(i);
  write_bit(out,cond(i));
  write_data(out,0,iter);
  write_data(out,1,val);
  i = next(i);
}
\end{lstlisting}
\end{minipage}
\vspace{-12pt}
\caption{Loop templates: (a) Regular, (b) Vectorizable}
\vspace{-10pt}
\label{fig:loop}
\end{figure}

Finally, Reffine lowers the generated imperative loops to LLVM IR and uses LLVM's JIT compiler~\cite{llvm} to produce hardware-specialized executable code. Our current implementation stores fields using Apache Arrow arrays~\cite{arrow}, providing a framework-agnostic columnar format that enables zero-copy interoperability with other analytics engines. Reffine also exploits Arrow's contiguous column buffers and bitmap-based null tracking to generate vectorization-friendly loops: the compiler rewrites each loop to operate directly on column buffers while recording iterator conditions in the corresponding bitmap buffer. As illustrated in Figure~\ref{fig:loop}b, this transformation produces a tight, column-oriented loop structure that LLVM can automatically vectorize when the loop bounds and control flow permit.

The generated loops are wrapped in functions that operate on input and output Arrow arrays together with boundary constraints. At runtime, Reffine assigns each worker thread a non-overlapping range ($[start, end)$) of the iteration space and invokes the compiled function in parallel. Threads read from shared input arrays while writing to disjoint regions of the output arrays, enabling efficient multi-core execution.

%% file: sections/60_eval.tex
\section{Evaluation}
\noindent\textbf{Benchmarks:} We prepare three sets of benchmarks. (1) A micro-benchmark consisting of four commonly used primitive operations like select, sum, inner, and outer join (2) seven relational queries from TPC-H~\cite{tpch} benchmark with different compute characteristics, and finally (3) two data analytics applications from streaming analytics, namely Trading and Normalize, and two from graph analytics, namely PageRank and Triangle counting, taken from prior works~\cite{weld,split,graphmat,tilt}. PageRank and Trading benchmarks are similar to the examples in Figure~\ref{fig:eg}. Normalize computes normalized values in a data stream over every $100$-day window. Triangle counting counts the triangles in the input graph. For the micro-benchmarks and streaming analytics benchmarks, we use a synthetic dataset with $100$ million entries each with a random timestamp and a floating point value. For the TPC-H benchmark, we generate the dataset using a scale factor of $10$. For the graph analytics, we use the SNAP Twitter graph~\cite{snap} containing $81,306$ nodes and more than $1.7$ million edges.

\noindent \textbf{Baselines:} We compare Reffine against five widely used and high performance data analytics systems: (1) Pandas~\cite{pandas}, one of the most widely used libraries for DataFrame analytics; (2) Polars~\cite{polars}, a high-performance multi-core DataFrame engine written in Rust; (3) DuckDB~\cite{duckdb}, a state-of-the-art in-process analytical database; (4) Umbra~\cite{umbra}, a state-of-the-art compiler-based relational database with high in-memory performance; and (5) NetworkX~\cite{netx}, a widely used library for graph analytics. Together, these systems provide representative and competitive baselines across the different data models and computational patterns targeted by Reffine.

\noindent \textbf{Experimental setup: } All the experiments are conducted on an AMD EPYC 7371 $16$-core processor ($32$-core with hyper-threading) and $128$ GB DRAM. Since different engines use different internal memory formats, we exclude data-loading time from our measurements to ensure a fair comparison. Instead, we report only the compute performance after the entire input dataset is loaded into memory. All the performance measurements reported are the median of five runs.

\subsection{Performance on Primitive Operations}
We evaluate the single-thread performance of Reffine on common primitive operations, such as select, sum, inner join, and outer join, using the synthetic dataset. The select operation scans the dataset and filters out $\sim50\%$ of the entries, while aggregate sums up all the values. The inner and outer join operations merge two synthetic datasets.

We compare Reffine’s performance against Pandas, Polars, and DuckDB. Figure~\ref{fig:micro} reports the normalized execution times for all four systems, using Pandas as the base. As shown, Reffine significantly outperforms libraries such as Pandas and Polars, with speedups ranging from $1.21-50\times$ over Pandas and $1.05-21\times$ over Polars. These results demonstrate that Reffine can generate highly efficient code even for primitive operations that are often hand-optimized in specialized engines. Against DuckDB, the state-of-the-art in-memory relational database, Reffine delivers comparable performance. For select and sum, DuckDB slightly outperforms Reffine, which we attribute to DuckDB’s more advanced SIMD optimizations. For inner and outer join operations, however, Reffine-generated code outperforms DuckDB’s implementation by up to $1.33\times$, highlighting Reffine’s ability to efficiently generate single-threaded co-iterating loops across two Fields to perform simple join operations.

\subsection{End-to-End Performance}
In this section, we evaluate Reffine's multi-core performance across a broad range of real-world data analytics applications using two benchmark suites. First, we use the industry-standard TPC-H decision support benchmark. Second, we evaluate Reffine on representative applications from streaming analytics and graph analytics. The results of these evaluations are presented below.

\subsubsection{TPC-H benchmark performance}\label{sec:tpch}
\begin{figure*}[h]%
    \centering
    \subfloat{
        \includegraphics[width=0.38\textwidth]{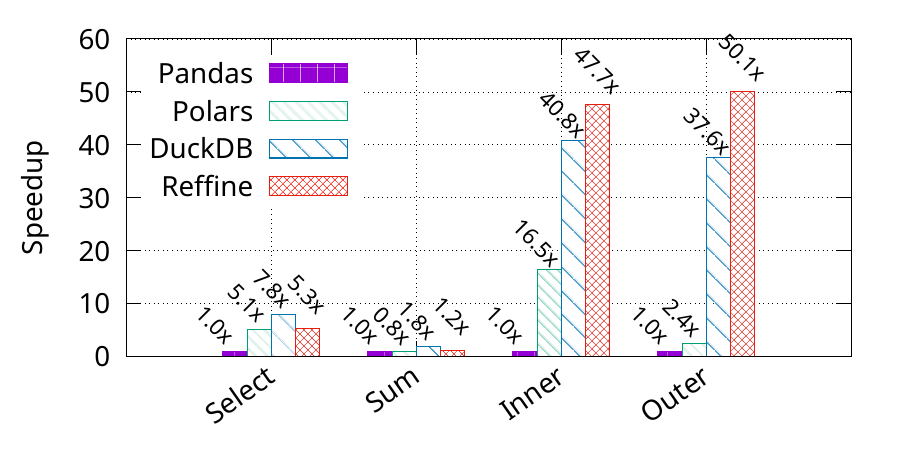}
        \label{fig:micro}
    }
    \subfloat{
        \includegraphics[width=0.57\textwidth]{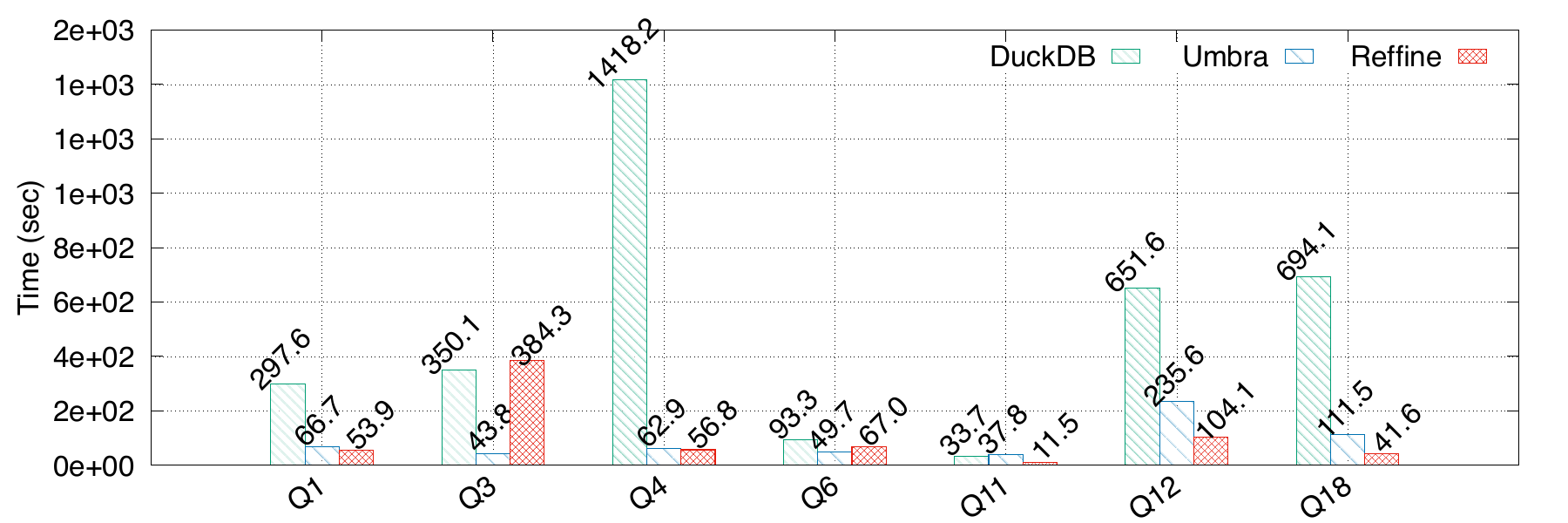}
        \label{fig:tpch}
    }
    \vspace{-10pt}
    \caption{Performance of Reffine on (a) primitive relational operations (b) TPC-H benchmark queries}%
\end{figure*}

We select seven queries from TPC-H with diverse computational characteristics and compare the multi-core performance of Reffine against DuckDB~\cite{duckdb} and Umbra~\cite{umbra}. All three systems are running with $16$-thread parallelism.

Figure~\ref{fig:tpch} presents the execution time of the seven TPC-H queries. Reffine outperforms DuckDB on six of the seven queries, with speedups of up to $24.9\times$. The largest gains occur on Queries 4 and 18, where Reffine achieves $24.9\times$ and $16.7\times$ speedups, respectively. Both queries contain expensive nested subqueries that are difficult for conventional relational engines to optimize effectively, as discussed in Section~\ref{sec:mot}. Reffine instead was able to lower the query into fused loops, allowing intermediate values to remain within the generated computation and exposing optimization and parallelization opportunities across operator boundaries.

Query 3 is the only exception, where Reffine is about $10\%$ slower than DuckDB. This query performs a three-way inner join with filtering and a grouped aggregation. We find that DuckDB’s advantage comes from its ability to choose an optimal join order based on the data distribution and cache certain intermediate results, whereas Reffine aggressively fuses the entire query into a single operation. We argue that these types of query-planning optimizations are orthogonal to Reffine’s design and such higher-level optimizations are better implemented above Reffine IR and then lowered into it. This allows combining the strengths of traditional query planning with Reffine’s efficient code-generation and execution model.

The comparison with Umbra is more competitive, as Umbra itself employs compilation-based query execution and already eliminates much of the interpretation and operator-dispatch overhead of traditional analytical engines. Nevertheless, Reffine outperforms Umbra on five of the seven queries, achieving speedups ranging from $1.1\times$ to $3.2\times$. These gains highlight the additional optimization opportunities exposed by Reffine's sparse representation and whole-computation analysis. In particular, Reffine can fuse computation across the entire query, whereas Umbra's fusion is primarily confined to individual pipelines and cannot cross blocking operators such as joins and aggregations~\cite{neumann,umbra}. Query 6, which consists primarily of a scan, filtering, and aggregation, is well-suited to Umbra's pipeline-oriented execution and leaves relatively little opportunity for Reffine to benefit from more aggressive cross-operator fusion; consequently, Umbra outperforms Reffine by $1.3\times$. Similarly, on Query 3, Umbra benefits from cost-based join planning, while Reffine is sensitive to the join order encoded in the input expression. Overall, these results show that although Umbra outperforms Reffine on two queries, Reffine achieves speedups of up to $3.2\times$ on the other five, demonstrating that it is competitive with a state-of-the-art compiled relational engine while providing a substantially more general programming abstraction.

\subsubsection{Performance on streaming and graph analytics}
In this section, we demonstrate how effectively Reffine accelerates non-relational data analytics workloads. To do so, we evaluate Reffine on two benchmark applications from streaming analytics against Polars and two from graph analytics against NetworkX Python library~\cite{netx}. We choose Polars over streaming engines like Trill~\cite{trill} because it delivers better performance on these benchmarks due to lower framework overhead. Figure~\ref{fig:nontpc} shows the execution time measured on all four benchmarks.

On streaming benchmark, Reffine achieves a $4.2\times$ speedup on Trading and a $32.4\times$ speedup on Normalize benchmarks respectively against Polars. The substantial gain on Normalize stems from Reffine’s ability to efficiently process temporal window operations using range-based indexing on fields. Polars, by contrast, must perform group-by aggregations to emulate the same behavior, resulting in higher overhead. On the PageRank and Triangle counting benchmarks, Reffine outperforms the NetworkX implementations by $2.5\times$ and $93.4\times$, respectively. For PageRank, NetworkX delegates computation to SciPy’s sparse matrix representation to avoid redundant work on the sparse graph and to leverage SciPy’s~\cite{scipy} optimized SpMV kernels. Despite this, Reffine still achieves a $2.5\times$ speedup, highlighting its efficiency in traversing sparse data structures. In contrast, the Triangle counting implementation in NetworkX is written mostly in Python, resulting in substantially higher overhead.

Overall, the above results show that Reffine is able to achieve high performance in a broad range of data analytics applications, often outperforming state-of-the-art specialized data analytics engines.

\subsection{Sensitivity Study}
\begin{figure*}[]%
    \centering
    \subfloat{
        \includegraphics[width=0.35\textwidth]{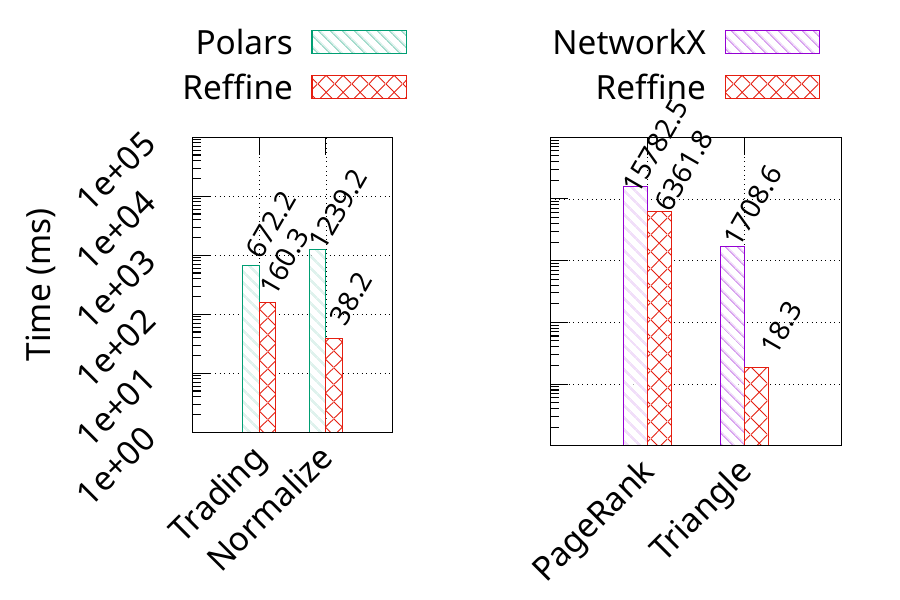}
        \label{fig:nontpc}
    }
    \subfloat{
        \includegraphics[width=0.30\textwidth]{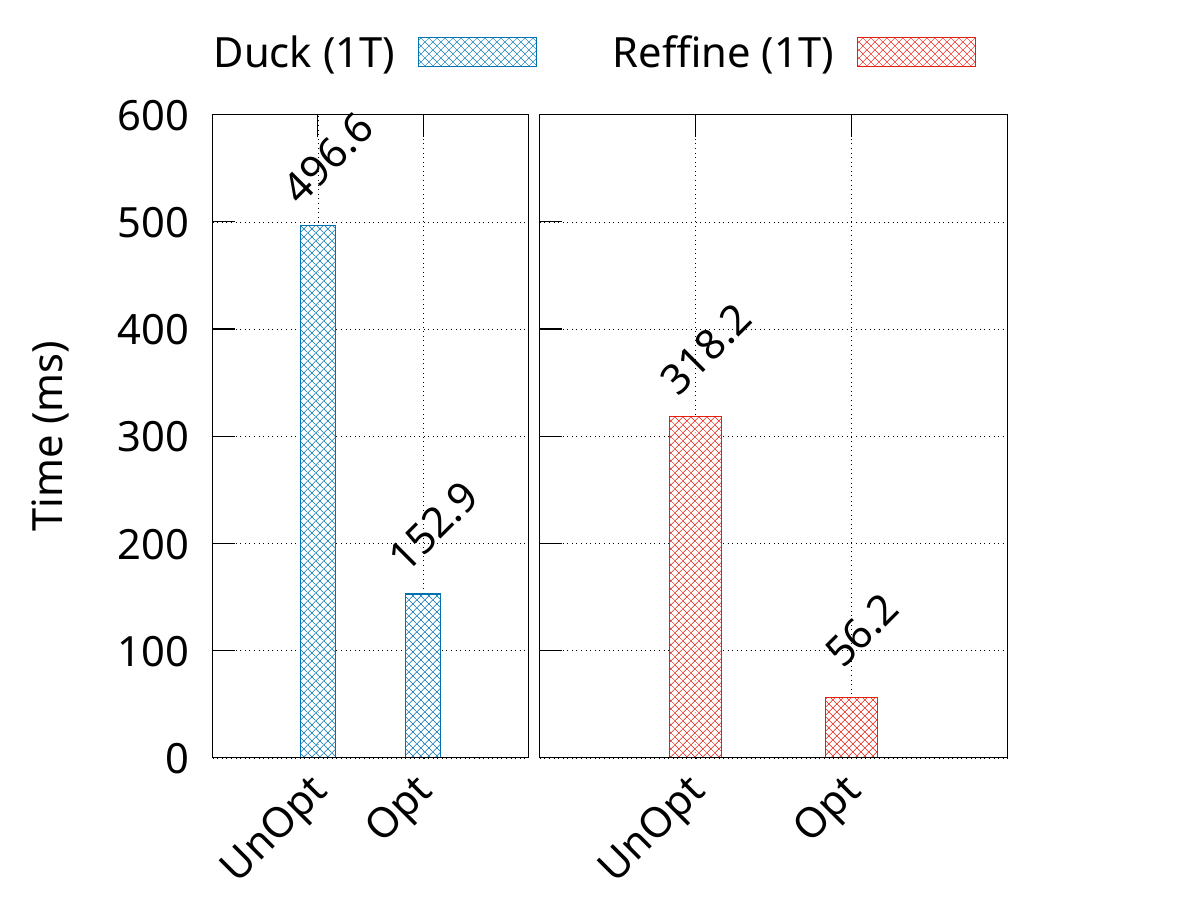}
        \label{fig:fuse}
    }
    \subfloat{
        \includegraphics[width=0.27\textwidth]{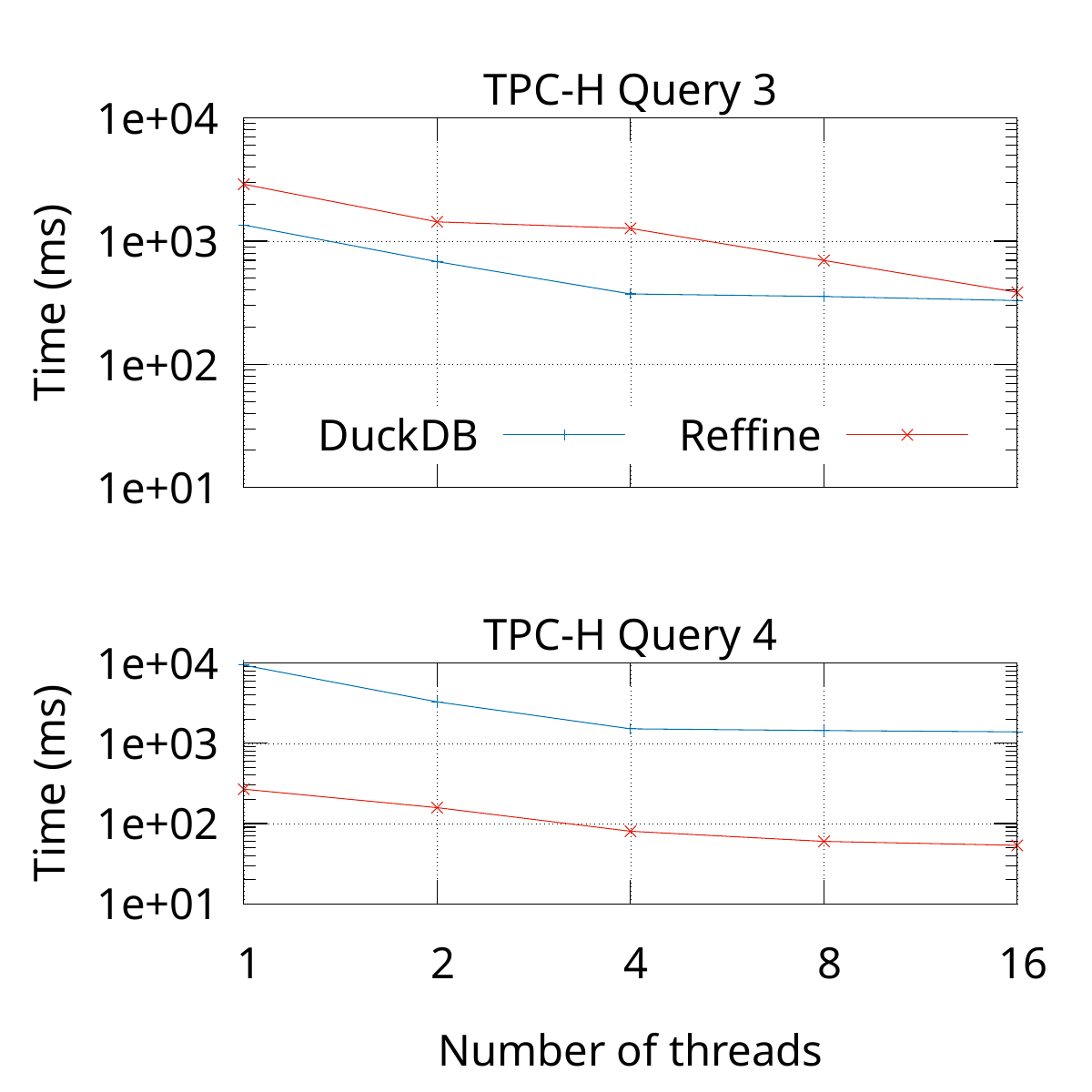}
        \label{fig:scale}
    }
    \vspace{-5pt}
    \caption{Execution time (a) on streaming and graph analytics (b) with and without operator fusion (c) parallel execution}%
    \vspace{-5pt}
\end{figure*}

We evaluate the benefits of the design decisions made in Reffine and how they contribute to its overall performance.

\noindent\textbf{Effectiveness of operator fusion}
We evaluate the effectiveness of Reffine’s fusion optimization by measuring the single-thread execution time of the example query in Figure~\ref{fig:eg}a before and after applying the fusion transformations described in Section~\ref{sec:pass}. We use single-thread performance here to avoid the effects of query parallelization in the measurements. We compare these results with DuckDB’s execution of both the un-optimized (Figure~\ref{fig:eg}a) and manually optimized (Listing~\ref{lst:tcphopt}) versions of the query. As shown in Figure~\ref{fig:fuse}, rewriting the SQL query in DuckDB yields a single-thread performance improvement of $3.2\times$ illustrating that even mature relational databases with advanced optimizers cannot automatically discover such transformations. In contrast, the un-optimized Reffine version already outperforms the un-optimized DuckDB by $1.5\times$, and applying fusion optimization further increases this speedup to $8.8\times$ ($2.7\times$ compared to the optimized version in DuckDB). This gain primarily comes from reducing intermediate materialization and improving data locality during execution. These findings show that Reffine’s greedy fusion strategy enables optimizations that other engines fail to exploit.

\noindent\textbf{Scalability}
To analyze the scalability of Reffine’s operation-level parallelization, we select two TPC-H queries: Query 3 and Query 4. Query 3 exhibits the worst speedup relative to DuckDB, while Query 4 exhibits the best. For both queries, we vary the number of threads from $1$ to $16$ and measure the execution time. Figure~\ref{fig:scale} reports the execution time across these concurrency levels. For Query 3, Reffine’s single-thread performance is $2.1\times$ slower than DuckDB. However, as we increase the number of threads, DuckDB’s performance improves only up to $4$ threads before plateauing, whereas Reffine continues to scale nearly linearly up to $16$ threads. As a result, despite its slower single-thread performance, Reffine ultimately matches DuckDB’s performance at higher thread counts. Query 4, on the other hand, exhibits a $35\times$ higher single-thread performance in Reffine compared to DuckDB. As discussed in Section~\ref{sec:tpch}, this improvement primarily comes from Reffine’s ability to aggressively fuse nested queries and generate highly efficient code. As concurrency increases, DuckDB again scales nearly linearly only up to $4$ threads before plateauing, whereas Reffine scales almost linearly up to $8$ threads, ultimately achieving a $24\times$ speedup. These results show that the combination of operator fusion and effective parallelization enables Reffine to deliver substantially higher in-memory performance than state-of-the-art data analytics engines.

%% file: sections/70_rw.tex
\section{Related Work}\label{sec:rw}

\noindent\textbf{Data analytics systems.}
The relational model~\cite{rel} provides a mathematically grounded abstraction for structured data, with a small set of algebraic operators that enables both expressive queries and efficient implementations. Modern analytical databases have built on this foundation with techniques such as columnar storage~\cite{monet,cstore}, multi-core execution, and compiler-based code generation~\cite{neumann}. As analytics workloads became more diverse, specialized abstractions and systems emerged for domains such as streaming~\cite{cql,millwheel,flink,dataflow,struct,trill}, numerical and DataFrame analytics~\cite{pandas,numpy}, and graph processing~\cite{pregel,graphx}. Reffine builds on established techniques such as columnar storage, compilation, and parallel execution, but introduces a more general programming model intended to unify these otherwise fragmented analytics domains.

\noindent\textbf{Sparse compilers.}
Sparse compilation was developed to efficiently execute computations over sparse data structures. TACO~\cite{taco} formalized sparse iteration theory and demonstrated how sparse tensor computations can be automatically lowered into efficient loops. Related efforts have applied sparse representations to relational and graph computations~\cite{indexstream,tra,graphmat}, but either focus on particular domains or remain primarily computational models. Reffine extends these ideas toward a general-purpose analytics engine capable of expressing relational, streaming, and graph computations.

Although Reffine could, in principle, be built on top of TACO, we chose to implement a compiler from scratch for several reasons. First, we found that TACO’s compute abstraction is insufficient to express the diverse set of data analytics workloads that we target. For example, it is difficult to express arbitrary filtering conditions or aggregate functions. Second, TACO’s compilation logic splits the iteration spaces into many smaller spaces, often producing numerous loops for a single operation. In contrast, Reffine generates a single loop per operator, yielding far simpler and more efficient code for data analytics workloads, better suited for parallelization and vectorization.

\noindent\textbf{Weld~\cite{weld}.}
Weld is the closest prior work to Reffine in its goal of providing a unified execution layer for diverse data analytics workloads. It enables cross-library optimization by representing computations using a common intermediate representation and applying techniques such as operator fusion and parallelization before generating code through LLVM. Weld IR exposes multiple data structures, primitive analytics operators, and parallel loop constructs, making it effective for dense array and map-reduce-style computations.

Reffine differs from Weld primarily in the design of its intermediate representation. Rather than exposing a collection of data structures and loop constructs, Reffine represents data using Fields and computation using Operators and Reductions, with sparse iteration providing the underlying execution model. This allows complex computations, including multi-table joins and temporal operations, to be expressed within a common abstraction while enabling fusion across operator boundaries and fine-grained dependency analysis for parallelization. A more detailed comparison between Reffine and Weld, including an empirical evaluation, is provided in the appendix~\ref{sec:weld}.

%% file: sections/80_con.tex
\section{Conclusion}
In this paper, we present Reffine, a sparse compiler-based in-memory analytics engine designed to deliver high performance across a broad spectrum of data analytics applications. Reffine introduces a novel intermediate representation and compiler design, grounded in relational algebra and sparse iteration theory, that enables optimizations such as operator fusion, data-parallel execution, and hardware-efficient code generation across diverse workloads. Our evaluation demonstrates that Reffine outperforms DuckDB by up to $24.9\times$ and the state-of-the-art compilation-based database Umbra by up to $3.2\times$ on TPC-H. Reffine also achieves average speedups of $18.3\times$ over Polars on streaming analytics and $47.9\times$ over NetworkX on graph analytics workloads. These results demonstrate that Reffine can provide a unified programming and execution model while remaining competitive with, and in many cases substantially outperforming, highly optimized domain-specific analytics systems.

%% file: sections/90_supp.tex
\section{Reffine v.s. Weld}\label{sec:weld}
We evaluate Reffine against Weld, an academic system with similar goals of designing a universal data analytics engine with compiler-based code generation and applying optimizations such as operator fusion, vectorization, and parallelization through its Weld IR and compiler backend. Unfortunately, the Weld code base is outdated and no longer actively maintained. Several operators required for TPC-H queries, such as multi-table joins, are unsupported or broken in the open-sourced version. Additionally, although Weld reportedly support multi-core execution, we were unable to get it to work in practice despite substantial tuning effort. As a result, we were unable to conduct a full evaluation of Weld on the TPC-H benchmark. Instead, we compare the single-thread performance of Weld and Reffine on the simplest query in the suite, Query 6.

\begin{figure}[h]
    \centering
    \includegraphics[width=0.30\textwidth]{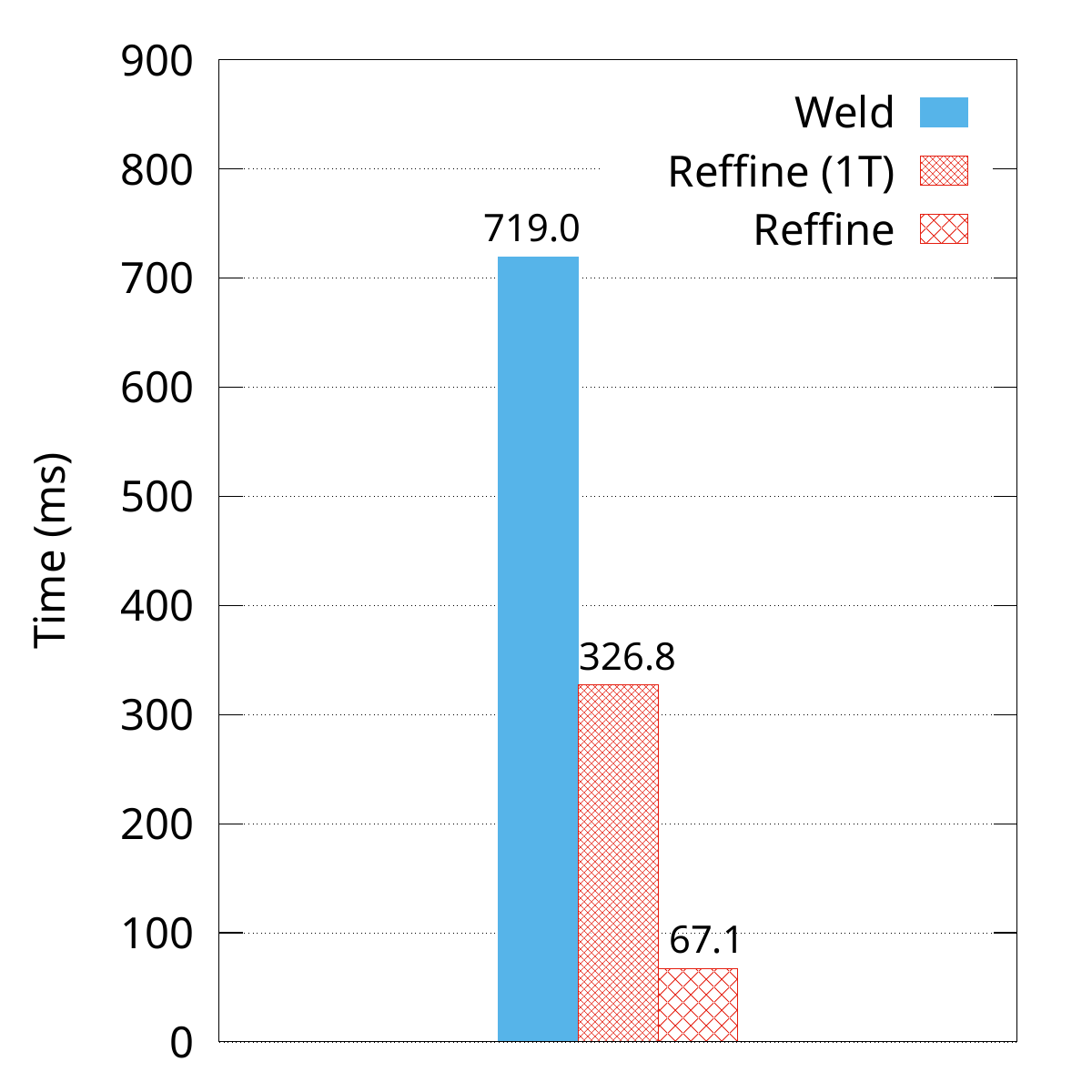}
    \caption{Performance of Reffine against Weld}
    \label{fig:weld}
\end{figure}

Figure~\ref{fig:weld} reports the single-thread performance of Weld alongside both the single-thread and $16$-threads performance of Reffine. As shown, single-threaded Reffine achieves a $2.2\times$ speedup over Weld, and with $16$ threads, Reffine reaches a $10.7\times$ speedup. These results indicate that Reffine’s optimizations and code-generation pipeline produce substantially more efficient execution than Weld. These results show that Reffine’s optimizations and code-generation pipeline yield substantially more efficient execution than Weld. Despite similar design goals, Reffine’s design and implementation differ significantly from Weld, which we believe contributes to its superior performance. We discuss these differences in more detail in Section~\ref{sec:rw}.

%% file: acmart.bib
@inproceedings{umbra,
  author       = {Thomas Neumann and
                  Michael J. Freitag},
  title        = {Umbra: {A} Disk-Based System with In-Memory Performance},
  booktitle    = {10th Conference on Innovative Data Systems Research, {CIDR} 2020,
                  Amsterdam, The Netherlands, January 12-15, 2020, Online Proceedings},
  publisher    = {www.cidrdb.org},
  year         = {2020},
  url          = {https://vldb.org/cidrdb/2020/umbra-a-disk-based-system-with-in-memory-performance.html},
  bibsource    = {dblp computer science bibliography, https://dblp.org}
}

@inproceedings{struct,
author = {Armbrust, Michael and Das, Tathagata and Torres, Joseph and Yavuz, Burak and Zhu, Shixiong and Xin, Reynold and Ghodsi, Ali and Stoica, Ion and Zaharia, Matei},
title = {Structured Streaming: A Declarative API for Real-Time Applications in Apache Spark},
year = {2018},
isbn = {9781450347037},
publisher = {Association for Computing Machinery},
address = {New York, NY, USA},
url = {https://doi.org/10.1145/3183713.3190664},
doi = {10.1145/3183713.3190664},
booktitle = {Proceedings of the 2018 International Conference on Management of Data},
pages = {601–613},
numpages = {13},
location = {Houston, TX, USA},
series = {SIGMOD '18}
}

@inproceedings{spark,
author = {Zaharia, Matei and Das, Tathagata and Li, Haoyuan and Hunter, Timothy and Shenker, Scott and Stoica, Ion},
title = {Discretized Streams: Fault-Tolerant Streaming Computation at Scale},
year = {2013},
isbn = {9781450323888},
publisher = {Association for Computing Machinery},
address = {New York, NY, USA},
url = {https://doi.org/10.1145/2517349.2522737},
doi = {10.1145/2517349.2522737},
booktitle = {Proceedings of the Twenty-Fourth ACM Symposium on Operating Systems Principles},
pages = {423–438},
numpages = {16},
location = {Farminton, Pennsylvania},
series = {SOSP '13}
}

@article{flink,
  author    = {Paris Carbone and
               Asterios Katsifodimos and
               Stephan Ewen and
               Volker Markl and
               Seif Haridi and
               Kostas Tzoumas},
  title     = {Apache Flink{\texttrademark}: Stream and Batch Processing in a Single
               Engine},
  journal   = {{IEEE} Data Eng. Bull.},
  volume    = {38},
  number    = {4},
  pages     = {28--38},
  year      = {2015},
  url       = {http://sites.computer.org/debull/A15dec/p28.pdf},
  bibsource = {dblp computer science bibliography, https://dblp.org}
}

@InProceedings{trill,
author = {Chandramouli, Badrish and Goldstein, Jonathan and Barnett, Mike and DeLine, Robert and Fisher, Danyel and Platt, John and Terwilliger, James and Wernsing, John and DeLIne, Robert},
title = {Trill: A High-Performance Incremental Query Processor for Diverse Analytics},
year = {2015},
month = {August},
publisher = {VLDB - Very Large Data Bases},
url = {https://www.microsoft.com/en-us/research/publication/trill-a-high-performance-incremental-query-processor-for-diverse-analytics/},
}

@article{dataflow,
    title	= {The Dataflow Model: A Practical Approach to Balancing Correctness, Latency, and Cost in Massive-Scale, Unbounded, Out-of-Order Data Processing},
    author	= {Tyler Akidau and Robert Bradshaw and Craig Chambers and Slava Chernyak and Rafael J. Fernández-Moctezuma and Reuven Lax and Sam McVeety and Daniel Mills and Frances Perry and Eric Schmidt and Sam Whittle},
    year	= {2015},
    journal	= {Proceedings of the VLDB Endowment},
    pages	= {1792-1803},
    volume	= {8}
}

@inproceedings{z3,
author = {De Moura, Leonardo and Bj\o{}rner, Nikolaj},
title = {Z3: an efficient SMT solver},
year = {2008},
isbn = {3540787992},
publisher = {Springer-Verlag},
address = {Berlin, Heidelberg},
booktitle = {Proceedings of the Theory and Practice of Software, 14th International Conference on Tools and Algorithms for the Construction and Analysis of Systems},
pages = {337–340},
numpages = {4},
location = {Budapest, Hungary},
series = {TACAS'08/ETAPS'08}
}

@inproceedings{llvm,
author = {Lattner, Chris and Adve, Vikram},
title = {LLVM: A Compilation Framework for Lifelong Program Analysis and Transformation},
year = {2004},
isbn = {0769521029},
publisher = {IEEE Computer Society},
address = {USA},
booktitle = {Proceedings of the International Symposium on Code Generation and Optimization: Feedback-Directed and Runtime Optimization},
pages = {75},
location = {Palo Alto, California},
series = {CGO '04}
}

@article{neumann,
author = {Neumann, Thomas},
title = {Efficiently Compiling Efficient Query Plans for Modern Hardware},
year = {2011},
issue_date = {June 2011},
publisher = {VLDB Endowment},
volume = {4},
number = {9},
issn = {2150-8097},
url = {https://doi.org/10.14778/2002938.2002940},
doi = {10.14778/2002938.2002940},
journal = {Proc. VLDB Endow.},
month = {jun},
pages = {539–550},
numpages = {12}
}

@inproceedings{lightsaber,
author = {Theodorakis, Georgios and Koliousis, Alexandros and Pietzuch, Peter and Pirk, Holger},
title = {LightSaber: Efficient Window Aggregation on Multi-Core Processors},
year = {2020},
isbn = {9781450367356},
publisher = {Association for Computing Machinery},
address = {New York, NY, USA},
url = {https://doi.org/10.1145/3318464.3389753},
doi = {10.1145/3318464.3389753},
booktitle = {Proceedings of the 2020 ACM SIGMOD International Conference on Management of Data},
pages = {2505–2521},
numpages = {17},
location = {Portland, OR, USA},
series = {SIGMOD '20}
}

@inproceedings {terse,
author = {Gennady Pekhimenko and Chuanxiong Guo and Myeongjae Jeon and Peng Huang and Lidong Zhou},
title = {{TerseCades}: Efficient Data Compression in Stream Processing},
booktitle = {2018 USENIX Annual Technical Conference (USENIX ATC 18)},
year = {2018},
isbn = {978-1-939133-01-4},
address = {Boston, MA},
pages = {307--320},
url = {https://www.usenix.org/conference/atc18/presentation/pekhimenko},
publisher = {USENIX Association},
month = jul,
}

@inproceedings{storm,
    author = {Veen, Jan Sipke van der and Waaij, Bram van der and Lazovik, Elena and Wijbrandi, Wilco and Meijer, Robert J.},
    title = {Dynamically Scaling Apache Storm for the Analysis of Streaming Data},
    year = {2015},
    isbn = {9781479981281},
    publisher = {IEEE Computer Society},
    address = {USA},
    url = {https://doi.org/10.1109/BigDataService.2015.56},
    doi = {10.1109/BigDataService.2015.56},
    booktitle = {Proceedings of the 2015 IEEE First International Conference on Big Data Computing Service and Applications},
    pages = {154–161},
    numpages = {8},
    series = {BIGDATASERVICE ’15}
}

@inproceedings{millwheel,
    title	= {MillWheel: Fault-Tolerant Stream Processing at Internet Scale},
    author	= {Tyler Akidau and Alex Balikov and Kaya Bekiroglu and Slava Chernyak and Josh Haberman and Reuven Lax and Sam McVeety and Daniel Mills and Paul Nordstrom and Sam Whittle},
    year	= {2013},
    booktitle	= {Very Large Data Bases},
    pages	= {734--746}
}

@inproceedings{streamboxhbm,
    author = {Miao, Hongyu and Jeon, Myeongjae and Pekhimenko, Gennady and McKinley, Kathryn S. and Lin, Felix Xiaozhu},
    title = {StreamBox-HBM: Stream Analytics on High Bandwidth Hybrid Memory},
    year = {2019},
    isbn = {9781450362405},
    publisher = {Association for Computing Machinery},
    address = {New York, NY, USA},
    url = {https://doi.org/10.1145/3297858.3304031},
    doi = {10.1145/3297858.3304031},
    booktitle = {Proceedings of the Twenty-Fourth International Conference on Architectural Support for Programming Languages and Operating Systems},
    pages = {167–181},
    numpages = {15},
    location = {Providence, RI, USA},
    series = {ASPLOS ’19}
}

@inproceedings{grizzly,
author = {Grulich, Philipp M. and Sebastian, Bre\ss{} and Zeuch, Steffen and Traub, Jonas and Bleichert, Janis von and Chen, Zongxiong and Rabl, Tilmann and Markl, Volker},
title = {Grizzly: Efficient Stream Processing Through Adaptive Query Compilation},
year = {2020},
isbn = {9781450367356},
publisher = {Association for Computing Machinery},
address = {New York, NY, USA},
url = {https://doi.org/10.1145/3318464.3389739},
doi = {10.1145/3318464.3389739},
booktitle = {Proceedings of the 2020 ACM SIGMOD International Conference on Management of Data},
pages = {2487–2503},
numpages = {17},
location = {Portland, OR, USA},
series = {SIGMOD '20}
}

@inproceedings{stream,
author = {Babcock, Brian and Babu, Shivnath and Datar, Mayur and Motwani, Rajeev and Widom, Jennifer},
title = {Models and Issues in Data Stream Systems},
year = {2002},
isbn = {1581135076},
publisher = {Association for Computing Machinery},
address = {New York, NY, USA},
url = {https://doi.org/10.1145/543613.543615},
doi = {10.1145/543613.543615},
booktitle = {Proceedings of the Twenty-First ACM SIGMOD-SIGACT-SIGART Symposium on Principles of Database Systems},
pages = {1–16},
numpages = {16},
location = {Madison, Wisconsin},
series = {PODS '02}
}

@article{cql,
author = {Arasu, Arvind and Babu, Shivnath and Widom, Jennifer},
title = {The CQL Continuous Query Language: Semantic Foundations and Query Execution},
year = {2006},
issue_date = {June 2006},
publisher = {Springer-Verlag},
address = {Berlin, Heidelberg},
volume = {15},
number = {2},
issn = {1066-8888},
url = {https://doi.org/10.1007/s00778-004-0147-z},
doi = {10.1007/s00778-004-0147-z},
journal = {The VLDB Journal},
month = {jun},
pages = {121–142},
numpages = {22}
}

@inproceedings{duckdb,
author = {Raasveldt, Mark and M\"{u}hleisen, Hannes},
title = {DuckDB: an Embeddable Analytical Database},
year = {2019},
isbn = {9781450356435},
publisher = {Association for Computing Machinery},
address = {New York, NY, USA},
url = {https://doi.org/10.1145/3299869.3320212},
doi = {10.1145/3299869.3320212},
booktitle = {Proceedings of the 2019 International Conference on Management of Data},
pages = {1981–1984},
numpages = {4},
location = {Amsterdam, Netherlands},
series = {SIGMOD '19}
}

@article{clickhouse,
author = {Schulze, Robert and Schreiber, Tom and Yatsishin, Ilya and Dahimene, Ryadh and Milovidov, Alexey},
title = {ClickHouse - Lightning Fast Analytics for Everyone},
year = {2024},
issue_date = {August 2024},
publisher = {VLDB Endowment},
volume = {17},
number = {12},
issn = {2150-8097},
url = {https://doi.org/10.14778/3685800.3685802},
doi = {10.14778/3685800.3685802},
journal = {Proc. VLDB Endow.},
month = aug,
pages = {3731–3744},
numpages = {14}
}

@article{postgresql,
author = {Stonebraker, M. and Rowe, L. A. and Hirohama, M.},
title = {The Implementation of POSTGRES},
year = {1990},
issue_date = {March 1990},
publisher = {IEEE Educational Activities Department},
address = {USA},
volume = {2},
number = {1},
issn = {1041-4347},
url = {https://doi.org/10.1109/69.50912},
doi = {10.1109/69.50912},
journal = {IEEE Trans. on Knowl. and Data Eng.},
month = mar,
pages = {125–142},
numpages = {18}
}

@Article{numpy,
 title         = {Array programming with {NumPy}},
 author        = {Charles R. Harris and K. Jarrod Millman and St{\'{e}}fan J.
                 van der Walt and Ralf Gommers and Pauli Virtanen and David
                 Cournapeau and Eric Wieser and Julian Taylor and Sebastian
                 Berg and Nathaniel J. Smith and Robert Kern and Matti Picus
                 and Stephan Hoyer and Marten H. van Kerkwijk and Matthew
                 Brett and Allan Haldane and Jaime Fern{\'{a}}ndez del
                 R{\'{i}}o and Mark Wiebe and Pearu Peterson and Pierre
                 G{\'{e}}rard-Marchant and Kevin Sheppard and Tyler Reddy and
                 Warren Weckesser and Hameer Abbasi and Christoph Gohlke and
                 Travis E. Oliphant},
 year          = {2020},
 month         = sep,
 journal       = {Nature},
 volume        = {585},
 number        = {7825},
 pages         = {357--362},
 doi           = {10.1038/s41586-020-2649-2},
 publisher     = {Springer Science and Business Media {LLC}},
 url           = {https://doi.org/10.1038/s41586-020-2649-2}
}

@ARTICLE{scipy,
  author  = {Virtanen, Pauli and Gommers, Ralf and Oliphant, Travis E. and
            Haberland, Matt and Reddy, Tyler and Cournapeau, David and
            Burovski, Evgeni and Peterson, Pearu and Weckesser, Warren and
            Bright, Jonathan and {van der Walt}, St{\'e}fan J. and
            Brett, Matthew and Wilson, Joshua and Millman, K. Jarrod and
            Mayorov, Nikolay and Nelson, Andrew R. J. and Jones, Eric and
            Kern, Robert and Larson, Eric and Carey, C J and
            Polat, {\.I}lhan and Feng, Yu and Moore, Eric W. and
            {VanderPlas}, Jake and Laxalde, Denis and Perktold, Josef and
            Cimrman, Robert and Henriksen, Ian and Quintero, E. A. and
            Harris, Charles R. and Archibald, Anne M. and
            Ribeiro, Ant{\^o}nio H. and Pedregosa, Fabian and
            {van Mulbregt}, Paul and {SciPy 1.0 Contributors}},
  title   = {{{SciPy} 1.0: Fundamental Algorithms for Scientific
            Computing in Python}},
  journal = {Nature Methods},
  year    = {2020},
  volume  = {17},
  pages   = {261--272},
  adsurl  = {https://rdcu.be/b08Wh},
  doi     = {10.1038/s41592-019-0686-2},
}

@software{pandas,
    author       = {The pandas development team},
    title        = {pandas-dev/pandas: Pandas},
    month        = feb,
    year         = 2020,
    publisher    = {Zenodo},
    version      = {latest},
    doi          = {10.5281/zenodo.3509134},
    url          = {https://doi.org/10.5281/zenodo.3509134}
}

@article{weld,
author = {Palkar, Shoumik and Thomas, James and Narayanan, Deepak and Thaker, Pratiksha and Palamuttam, Rahul and Negi, Parimajan and Shanbhag, Anil and Schwarzkopf, Malte and Pirk, Holger and Amarasinghe, Saman and Madden, Samuel and Zaharia, Matei},
title = {Evaluating end-to-end optimization for data analytics applications in weld},
year = {2018},
issue_date = {May 2018},
publisher = {VLDB Endowment},
volume = {11},
number = {9},
issn = {2150-8097},
url = {https://doi.org/10.14778/3213880.3213890},
doi = {10.14778/3213880.3213890},
journal = {Proc. VLDB Endow.},
month = may,
pages = {1002–1015},
numpages = {14}
}

@misc{cidr,
	author = {},
	title = {{W}eld: {A} common runtime for high performance data analytics --- dspace.mit.edu},
	howpublished = {\url{https://dspace.mit.edu/handle/1721.1/137425}},
	year = {},
	note = {[Accessed 11-12-2025]},
}

@inproceedings{split,
author = {Palkar, Shoumik and Zaharia, Matei},
title = {Optimizing data-intensive computations in existing libraries with split annotations},
year = {2019},
isbn = {9781450368735},
publisher = {Association for Computing Machinery},
address = {New York, NY, USA},
url = {https://doi.org/10.1145/3341301.3359652},
doi = {10.1145/3341301.3359652},
booktitle = {Proceedings of the 27th ACM Symposium on Operating Systems Principles},
pages = {291–305},
numpages = {15},
location = {Huntsville, Ontario, Canada},
series = {SOSP '19}
}

@PhDThesis{taco,
  title = "Sparse Tensor Algebra Compilation",
  author = "Fredrik Kjolstad",
  month = "Feb",
  year = "2020",
  url = "http://tensor-compiler.org/files/kjolstad-phd-thesis-taco-compiler.pdf",
  type = "Ph.D. Thesis",
  address = "Cambridge, MA",
  school = "Massachusetts Institute of Technology",
}

@article{indexstream,
author = {Kovach, Scott and Kolichala, Praneeth and Gu, Tiancheng and Kjolstad, Fredrik},
title = {Indexed Streams: A Formal Intermediate Representation for Fused Contraction Programs},
year = {2023},
issue_date = {June 2023},
publisher = {Association for Computing Machinery},
address = {New York, NY, USA},
volume = {7},
number = {PLDI},
url = {https://doi.org/10.1145/3591268},
doi = {10.1145/3591268},
journal = {Proc. ACM Program. Lang.},
month = jun,
articleno = {154},
numpages = {25}
}

@article{rel,
author = {Codd, E. F.},
title = {A relational model of data for large shared data banks},
year = {1970},
issue_date = {June 1970},
publisher = {Association for Computing Machinery},
address = {New York, NY, USA},
volume = {13},
number = {6},
issn = {0001-0782},
url = {https://doi.org/10.1145/362384.362685},
doi = {10.1145/362384.362685},
journal = {Commun. ACM},
month = jun,
pages = {377–387},
numpages = {11}
}

@inproceedings{tilt,
author = {Jayarajan, Anand and Zhao, Wei and Sun, Yudi and Pekhimenko, Gennady},
title = {TiLT: A Time-Centric Approach for Stream Query Optimization and Parallelization},
year = {2023},
isbn = {9781450399166},
publisher = {Association for Computing Machinery},
address = {New York, NY, USA},
url = {https://doi.org/10.1145/3575693.3575704},
doi = {10.1145/3575693.3575704},
booktitle = {Proceedings of the 28th ACM International Conference on Architectural Support for Programming Languages and Operating Systems, Volume 2},
pages = {818–832},
numpages = {15},
location = {Vancouver, BC, Canada},
series = {ASPLOS 2023}
}

@inproceedings{tpch,
author = {Boncz, Peter and Neumann, Thomas and Erling, Orri},
title = {TPC-H Analyzed: Hidden Messages and Lessons Learned from an Influential Benchmark},
year = {2013},
isbn = {9783319049359},
publisher = {Springer-Verlag},
address = {Berlin, Heidelberg},
url = {https://doi.org/10.1007/978-3-319-04936-6_5},
doi = {10.1007/978-3-319-04936-6_5},
booktitle = {Revised Selected Papers of the 5th TPC Technology Conference on Performance Characterization and Benchmarking - Volume 8391},
pages = {61–76},
numpages = {16}
}

@inproceedings{netx,
  address = {Pasadena, CA USA},
  author = {Hagberg, Aric A. and Schult, Daniel A. and Swart, Pieter J.},
  booktitle = {Proceedings of the 7th Python in Science Conference},
  editor = {Varoquaux, Ga\"el and Vaught, Travis and Millman, Jarrod},
  pages = {11 - 15},
  title = {Exploring Network Structure, Dynamics, and Function using NetworkX},
  url = {http://conference.scipy.org/proceedings/SciPy2008/paper_2/},
  year = 2008
}

@Manual{polars,
title = {polars: R Bindings for the 'polars' Rust Library},
author = {Tatsuya Shima and Etienne Bacher and {Authors of the
  dependency Rust crates}},
year = {2025},
url = {https://github.com/pola-rs/r-polars},
}

@inproceedings{sparksql,
author = {Armbrust, Michael and Xin, Reynold S. and Lian, Cheng and Huai, Yin and Liu, Davies and Bradley, Joseph K. and Meng, Xiangrui and Kaftan, Tomer and Franklin, Michael J. and Ghodsi, Ali and Zaharia, Matei},
title = {Spark SQL: Relational Data Processing in Spark},
year = {2015},
isbn = {9781450327589},
publisher = {Association for Computing Machinery},
address = {New York, NY, USA},
url = {https://doi.org/10.1145/2723372.2742797},
doi = {10.1145/2723372.2742797},
booktitle = {Proceedings of the 2015 ACM SIGMOD International Conference on Management of Data},
pages = {1383–1394},
numpages = {12},
location = {Melbourne, Victoria, Australia},
series = {SIGMOD '15}
}

@inproceedings{graphx,
author = {Xin, Reynold S. and Gonzalez, Joseph E. and Franklin, Michael J. and Stoica, Ion},
title = {GraphX: a resilient distributed graph system on Spark},
year = {2013},
isbn = {9781450321884},
publisher = {Association for Computing Machinery},
address = {New York, NY, USA},
url = {https://doi.org/10.1145/2484425.2484427},
doi = {10.1145/2484425.2484427},
booktitle = {First International Workshop on Graph Data Management Experiences and Systems},
articleno = {2},
numpages = {6},
location = {New York, New York},
series = {GRADES '13}
}

@inproceedings{optbig,
author = {Schlaipfer, Matthias and Rajan, Kaushik and Lal, Akash and Samak, Malavika},
title = {Optimizing Big-Data Queries Using Program Synthesis},
year = {2017},
isbn = {9781450350853},
publisher = {Association for Computing Machinery},
address = {New York, NY, USA},
url = {https://doi.org/10.1145/3132747.3132773},
doi = {10.1145/3132747.3132773},
booktitle = {Proceedings of the 26th Symposium on Operating Systems Principles},
pages = {631–646},
numpages = {16},
location = {Shanghai, China},
series = {SOSP '17}
}

@article{graphmat,
author = {Sundaram, Narayanan and Satish, Nadathur and Patwary, Md Mostofa Ali and Dulloor, Subramanya R. and Anderson, Michael J. and Vadlamudi, Satya Gautam and Das, Dipankar and Dubey, Pradeep},
title = {GraphMat: high performance graph analytics made productive},
year = {2015},
issue_date = {July 2015},
publisher = {VLDB Endowment},
volume = {8},
number = {11},
issn = {2150-8097},
url = {https://doi.org/10.14778/2809974.2809983},
doi = {10.14778/2809974.2809983},
journal = {Proc. VLDB Endow.},
month = jul,
pages = {1214–1225},
numpages = {12}
}

@article{tra,
author = {Yuan, Binhang and Jankov, Dimitrije and Zou, Jia and Tang, Yuxin and Bourgeois, Daniel and Jermaine, Chris},
title = {Tensor relational algebra for distributed machine learning system design},
year = {2021},
issue_date = {April 2021},
publisher = {VLDB Endowment},
volume = {14},
number = {8},
issn = {2150-8097},
url = {https://doi.org/10.14778/3457390.3457399},
doi = {10.14778/3457390.3457399},
journal = {Proc. VLDB Endow.},
month = apr,
pages = {1338–1350},
numpages = {13}
}

@online{avx,
  author = {Intel},
  title = {Intel Advanced Vector Extensions 512},
  url = {https://www.intel.com/content/www/us/en/architecture-and-technology/avx-512-overview.html},
}

@inproceedings{drc,
author = {Lacroix, Michel and Pirotte, Alain},
title = {Domain-oriented relational languages},
year = {1977},
publisher = {VLDB Endowment},
booktitle = {Proceedings of the Third International Conference on Very Large Data Bases - Volume 3},
pages = {370–378},
numpages = {9},
location = {Tokyo, Japan},
series = {VLDB '77}
}

@Manual{dask,
  title = {Dask: Library for dynamic task scheduling},
  author = {{Dask Development Team}},
  year = {2016},
  url = {http://dask.pydata.org},
}

@INPROCEEDINGS{onesize,
  author={Stonebraker, M. and Cetintemel, U.},
  booktitle={21st International Conference on Data Engineering (ICDE'05)}, 
  title={"One size fits all": an idea whose time has come and gone}, 
  year={2005},
  volume={},
  number={},
  pages={2-11},
  doi={10.1109/ICDE.2005.1}}

@article{job,
author = {Leis, Viktor and Gubichev, Andrey and Mirchev, Atanas and Boncz, Peter and Kemper, Alfons and Neumann, Thomas},
title = {How good are query optimizers, really?},
year = {2015},
issue_date = {November 2015},
publisher = {VLDB Endowment},
volume = {9},
number = {3},
issn = {2150-8097},
url = {https://doi.org/10.14778/2850583.2850594},
doi = {10.14778/2850583.2850594},
journal = {Proc. VLDB Endow.},
month = nov,
pages = {204–215},
numpages = {12}
}

@misc{arrow,
  title = {Apache Arrow},
  howpublished = {\url{https://arrow.apache.org/}},
  note = {Accessed: 2026-09-10}
}

@inproceedings{snap,
author = {McAuley, Julian and Leskovec, Jure},
title = {Learning to discover social circles in ego networks},
year = {2012},
publisher = {Curran Associates Inc.},
address = {Red Hook, NY, USA},
booktitle = {Proceedings of the 26th International Conference on Neural Information Processing Systems - Volume 1},
pages = {539–547},
numpages = {9},
location = {Lake Tahoe, Nevada},
series = {NIPS'12}
}

@inproceedings{monet,
author = {Blockhaus, Paul and Broneske, David and Sch\"{a}ler, Martin and K\"{o}ppen, Veit and Saake, Gunter},
title = {Combining Two Worlds: MonetDB with Multi-Dimensional Index Structure Support to Efficiently Query Scientific Data},
year = {2020},
isbn = {9781450388146},
publisher = {Association for Computing Machinery},
address = {New York, NY, USA},
url = {https://doi.org/10.1145/3400903.3401691},
doi = {10.1145/3400903.3401691},
booktitle = {Proceedings of the 32nd International Conference on Scientific and Statistical Database Management},
articleno = {29},
numpages = {4},
location = {Vienna, Austria},
series = {SSDBM '20}
}

@inproceedings{cstore,
author = {Stonebraker, Mike and Abadi, Daniel J. and Batkin, Adam and Chen, Xuedong and Cherniack, Mitch and Ferreira, Miguel and Lau, Edmond and Lin, Amerson and Madden, Sam and O'Neil, Elizabeth and O'Neil, Pat and Rasin, Alex and Tran, Nga and Zdonik, Stan},
title = {C-store: a column-oriented DBMS},
year = {2005},
isbn = {1595931546},
publisher = {VLDB Endowment},
booktitle = {Proceedings of the 31st International Conference on Very Large Data Bases},
pages = {553–564},
numpages = {12},
location = {Trondheim, Norway},
series = {VLDB '05}
}

@inproceedings{pregel,
author = {Malewicz, Grzegorz and Austern, Matthew H. and Bik, Aart J.C and Dehnert, James C. and Horn, Ilan and Leiser, Naty and Czajkowski, Grzegorz},
title = {Pregel: a system for large-scale graph processing},
year = {2010},
isbn = {9781450300322},
publisher = {Association for Computing Machinery},
address = {New York, NY, USA},
url = {https://doi.org/10.1145/1807167.1807184},
doi = {10.1145/1807167.1807184},
booktitle = {Proceedings of the 2010 ACM SIGMOD International Conference on Management of Data},
pages = {135–146},
numpages = {12},
location = {Indianapolis, Indiana, USA},
series = {SIGMOD '10}
}
